\documentclass[twocolumn]{aastex631}

\usepackage{threeparttable}
\usepackage{multirow,xspace,upgreek}
\usepackage[normalem]{ ulem }
\usepackage{soul}
\usepackage{colortbl}
\usepackage{grffile, graphicx}
\usepackage[T1]{fontenc}
\usepackage{lmodern}
\usepackage{txfonts}
\newcommand{\mic}{\ensuremath{\upmu}m\xspace}
\newcommand{\Teff}{\ensuremath{\mathrm{T_{eff}}}\xspace}
\newcommand{\fsed}{\ensuremath{\mathrm{f_{sed}}}\xspace}
\newcommand{\Rjup}{\ensuremath{\mathrm{R_{Jup}}}\xspace}
\newcommand{\Mjup}{\ensuremath{\mathrm{M_{Jup}}}\xspace}
\newcommand{\Rlam}{\ensuremath{\mathrm{R_{\uplambda}}}\xspace}
\newcommand{\chr}{\ensuremath{\mathrm{\chi^{2}_{r}}}\xspace}

\usepackage{natbib}

\begin{document}

\title{\large The \textit{JWST} Early Release Science Program for Direct Observations of Exoplanetary Systems V: \\Do Self-Consistent Atmospheric Models Represent JWST Spectra? A Showcase With VHS~1256-1257~b.}


\author[0000-0003-0331-3654]{Simon Petrus}\thanks{Send manuscript correspondence to simon.petrus.pro@gmail.com}
\affiliation{N\'{u}cleo Milenio Formac\'{i}on Planetaria - NPF, Universidad de Valpara\'{i}so, Av. Gran Breta\~{n}a 1111, Valpara\'{i}so, Chile}
\affiliation{Millennium Nucleus on Young Exoplanets and their Moons (YEMS), Chile}
\affiliation{Departamento de F\'{i}sica, Universidad de Santiago de Chile, Av. Victor Jara 3659, Santiago, Chile}

\author[0000-0001-8818-1544]{Niall Whiteford}
\affiliation{Department of Astrophysics, American Museum of Natural History, Central Park West at 79th Street, New York, NY 10034, USA}

\author[0000-0001-8718-3732]{Polychronis Patapis}
\affiliation{Institute of Particle Physics and Astrophysics, ETH Zurich, Wolfgang-Pauli-Str. 27, 8093 Zurich, Switzerland}

\author[0000-0003-4614-7035]{Beth A. Biller}
\affiliation{Scottish Universities Physics Alliance, Institute for Astronomy, University of Edinburgh, Blackford Hill, Edinburgh EH9 3HJ, UK}
\affiliation{Centre for Exoplanet Science, University of Edinburgh, Edinburgh EH9 3HJ, UK}

\author[0000-0001-6098-3924]{Andrew Skemer} 
\affiliation{University of California Santa Cruz, Santa Cruz, CA, USA}

\author[0000-0001-8074-2562]{Sasha Hinkley}
\affiliation{University of Exeter, Astrophysics Group, Physics Building, Stocker Road, Exeter, EX4 4QL, UK}

\author[0000-0002-2011-4924]{Genaro Suárez}
\affiliation{Department of Astrophysics, American Museum of Natural History, Central Park West at 79th Street, New York, NY 10034, USA}

\author[0000-0002-6217-6867]{Paulina Palma-Bifani}
\affiliation{Laboratoire Lagrange, Université Cote d’Azur, CNRS, Observatoire de la Cote d’Azur, 06304 Nice, France}
\affiliation{LESIA, Observatoire de Paris, Univ PSL, CNRS, Sorbonne Univ, Univ de Paris, 5 place Jules Janssen, 92195 Meudon, France}

\author[0000-0002-4404-0456]{Caroline V. Morley}
\affiliation{University of Texas at Austin, 2515 Speedway, Austin, TX 78712}

\author[0000-0001-6172-3403]{Pascal Tremblin}
\affiliation{Maison de la Simulation, CEA, CNRS, Univ. Paris-Sud, UVSQ, Université Paris-Saclay, F-91191 Gif-sur-Yvette, France}

% ...

\begin{abstract}

The unprecedented medium-resolution (\Rlam~$\sim$1500-3500) near- and mid-infrared (1-18\mic) spectrum provided by JWST for the young (140~$\pm$20Myr) low-mass (12-20~\Mjup) L-T transition (L7) companion VHS~1256~b gives access to a catalogue of molecular absorptions. In this study, we present a comprehensive analysis of this dataset utilizing a forward modelling approach, applying our Bayesian framework, \texttt{ForMoSA}. We explore five distinct atmospheric models to assess their performance in estimating key atmospheric parameters: \Teff, log(g), [M/H], C/O, $\gamma$, \fsed, and R. Our findings reveal that each parameter's estimate is significantly influenced by factors such as the wavelength range considered and the model chosen for the fit. This is attributed to systematic errors in the models and their challenges in accurately replicating the complex atmospheric structure of VHS~1256~b, notably the complexity of its clouds and dust distribution. To propagate the impact of these systematic uncertainties on our atmospheric property estimates, we introduce innovative fitting methodologies based on independent fits performed on different spectral windows. We finally derived a \Teff consistent with the spectral type of the target, considering its young age, which is confirmed by our estimate of log(g). Despite the exceptional data quality, attaining robust estimates for chemical abundances [M/H] and C/O, often employed as indicators of formation history, remains challenging. Nevertheless, the pioneering case of JWST's data for VHS~1256~b has paved the way for future acquisitions of substellar spectra that will be systematically analyzed to directly compare the properties of these objects and correct the systematics in the models.

\end{abstract}


\section{Introduction}\label{sec:intro}
\textcolor{black}{
The diversity of exoplanets that have been discovered to date has reshaped our understanding of planetary systems and has raised new questions regarding their formation and evolution. Although different formation scenarios, such as core accretion \citep{Pollack96}, gravitational instability \citep{Boss00}, and gravo-turbulent fragmentation of molecular clouds \citep{Padoan05}, have been proposed, constraining the evolutionary history of observed planets remains challenging. Indeed, the initial physical and chemical conditions of the circumstellar disk in which they formed are no longer directly observable, and the dynamical evolution of the system (migration, ejection, or planetary capture) is unknown. However, formation models have identified key parameters in the atmosphere of young planets that can potentially serve as formation tracers, such as the metallicity [M/H] \citep{Ormel21}, the carbon-to-oxygen ratio C/O \citep{Oberg11}, and, more recently, the isotopic ratio of $^{12}$CO/$^{13}$CO \citep{Zhang21}. By detecting specific molecules in the atmospheres of transiting planets, it has become possible to estimate these observables \citep{Hoeijmakers19, Spake21}. However, this technique is limited to highly irradiated planets close to their host stars. Direct imaging allows for the characterization of the atmospheres of planets with larger separations, but due to the high contrast with their host star, only sufficiently bright companions can be accessed. 
Presently, there are fewer than thirty low-mass companions for which spectroscopic data have been obtained. Only a few of these observations have reached sufficient spectral resolution (\Rlam $>$ 1000) to attempt estimating chemical abundances in their atmospheres,  including $\rm \beta$~Pic~b \citep{Snellen14}, HD~106906~b \citep{Daemgen17}, 1RXS~J1609~b \citep{Lafreniere10}, Delorme~1(AB)~b \citep{Eriksson20, Betti22, Ringqvist23}, HR~8799~b,~c \citep{Konopacky13, Ruffio19, Wang21, Wang23}, $\rm \kappa$~And~b \citep{Wilcomb20}, TYC~8998~b \citep{Zhang21}, HIP~65426~b \citep{Petrus21}, AB~Pic~b \cite{Palma23}, and lastly VHS~1256~b \citep{Petrus23, Hoch22} which is the subject of this work.}

\textcolor{black}{The system VHS~J125601.92-125723.9 (hereafter VHS~1256) has been identified as a hierarchical system with the detection of the companion VHS~1256~b, orbiting at $\sim$ 8.06~$\pm$~0.03~arcsec (projected physical separation of 179~$\pm$~9~au) from the tight (0.1~as) M7.5+M7.5 binary VHS~1256~AB \citep{Gauza15, Stone16}. Its distance and age have been estimated to 21.14~$\pm$~0.22~pc \citep{Gaia21} and 140~$\pm$~20~Myr \citep{Dupuy23}, respectively.}

\textcolor{black}{\cite{Bowler20} and \cite{Zhou20} monitored the companion with HST and Spitzer, respectively, and revealed significant wavelength-dependent variability, corresponding to an estimated rotation period ranging between 21 and 24 hours. \cite{Zhou22} confirmed this hourly-period variability and further identified a higher temporal baseline variability with multi-epoch tracking over a two-years period. Both publications emphasized VHS~1256~b as one of the most variable substellar objects discovered this far (over 20\% between 1.1 and 1.7~\mic) and interpreted this variability as the signature of a complex and heterogeneous atmospheric structure (inhomogeneous clouds cover, heterogeneous temperature, etc.). These conclusions align with the results provided by \cite{Gauza15} who identified VHS~1256~b as an L7-type object, at the boundary of the L-T transition (between 1000-1400~K). Through this temperature range, planetary atmospheres are expected to undergo significant modifications to their cloud structure in terms of vertical distribution (condensation and/or sedimentation of FeH, TiO, VO, \citealt{Cushing08}), and/or horizontal distribution (appearance of holes in the cloud cover, \citealt{Burgasser02, Marley10}). 
Furthermore, by employing the methodology detailed in \cite{Allers13}, \cite{Gauza15} noted indicators of low surface gravity within the spectrum of this object, consistent with its young age. Notably, they observed a characteristic triangular H-band shape. This low gravity implied a reddening effect (see Figure 1 in \citealt{Miles22}) that provides additional evidence supporting the presence of cloudy layers.}

\textcolor{black}{Several studies have analyzed medium- and high-resolution spectra of VHS~1256~b to better understand this complexity. Using Keck/NIRSPEC data in the H band (\Rlam$\sim$ 25000) \cite{Bryan18} estimated a projected rotational velocity v.sin(i) of 13.5~$\pm$~4.1~km.s$^{-1}$  which was consistent with the rotational period estimated by \cite{Zhou20}. This v.sin(i) measurement also enabled \cite{Zhou20} to estimate that VHS~1256~b is very likely to be observed equatorially. Keck/NIRSPEC L-band spectroscopy ($\Delta\uplambda$~= 2.9 to 4.4~\mic; \Rlam $\sim$ 1300) enabled the detection of methane, which revealed non-equilibrium chemistry between CO and CH$_{4}$ \citep{Miles18}. \cite{Hoch22} compared medium-resolution Keck/OSIRIS spectroscopy in the K band (\Rlam $\sim$ 4000) with a custom grid of pre-computed synthetic spectra to derive a \Teff range of 1200–1300~K, a log(g) range of 3.25–3.75~dex, and to estimate a C/O of 0.590~$\pm$~0.354. These atmospheric parameter estimates were confirmed by \cite{Petrus23} who fit a medium-resolution spectrum from VLT/X-Shooter (\Rlam $\sim$ 8000) spanning simultaneously 1.10 to 2.48~\mic with the \texttt{ATMO} grid \citep{Tremblin15} to estimate \Teff~=~1380~$\pm$~54~K, log(g)~=~3.97~$\pm$~0.48~dex, [M/H]~=~0.21~$\pm$~0.29, and C/O~>~0.63. They concluded that while the estimates of [M/H] and C/O suggested supersolar values, indicating significant enrichment in solids during its formation, the hypothesis of a formation by fragmentation of the initial molecular clouds could not be dismissed due to the characteristics of the system's architecture (hierarchical system) and the lack of sufficient precision in estimates of these formation tracers.}

\textcolor{black}{More recently, VHS~1256~b has been observed with JWST, which combined medium-resolution spectroscopy from NIRSpec and MIRI \citep{Miles22} (see Section \ref{sec:data}). The spectra were integrated to obtain a robust estimate of the bolometric luminosity log(L$\rm _{bol}$/L$_{\odot}$)~=~-4.550~$\pm$~0.009. \cite{Dupuy23} combined this luminosity with their estimate of the age to derive the mass of VHS~1256~b using the hybrid cloudy evolutionary model of \cite{Saumon08}. They identified a bi-modal solution with a lower-mass regime at 12~$\pm$~0.1~\Mjup and a higher-mass regime at 16~$\pm$~1~\Mjup. Given this bimodality, we can confidently place the mass between 12-20~\Mjup, but, as we don't know whether this object is currently burning deuterium, we don't know which of the bimodal peaks it truly falls in. \cite{Dupuy23} also estimated a radius R of 1.30~\Rjup and 1.22~\Rjup, a \Teff of 1153~$\pm$~5~K and 1194~$\pm$~9~K, and a log(g) of 4.268~$\pm$~0.006~dex and 4.45~$\pm$~0.03~dex according to the lower-mass and higher-mass solution, respectively. \cite{Miles22} also compared this spectrum with a grid of synthetic spectra from \cite{Mukherjee23} to estimate a \Teff of 1100~K, a log(g) of 4.5~dex, a radius of 1.27~\Rjup, and to highlight a complex atmosphere with different cloud deck and varying regimes of vertical mixing. Additionally, the MIRI data enabled the detection of a strong silicate absorption at around 10~\mic, supporting the presence of silicate clouds in its atmosphere.}

\textcolor{black}{These previous studies indicate the usefulness of employing grids of pre-computed synthetic spectra for characterizing the atmosphere of VHS~1256~b. These grids are generated from atmospheric models that incorporate our current understanding of the physics and chemistry involved. However, it has been observed that these models sometimes encounter challenges in accurately reproducing certain spectral features and extended wavelength ranges. These systematic errors can be attributed to the use of strong assumptions when constraining the physical and chemical processes occurring in planetary atmospheres, such as cloud formation and evolution, chemical disequilibrium, vertical mixing, and dust sedimentation (see Section \ref{sec:mod}). The imposition of these assumptions leads to a limited number of free parameters (< 6), and the quality of the fit heavily relies on the quality of the models. The impacts of these systematic errors on the estimation of atmospheric properties have already been recognized for L and M spectral types \citep{Cushing08, Stephens09, Manjavacas14, Bonnefoy14, Lachapelle15, Bayo17, Petrus20, Suarez21, Petrus23, Lueber23, Palma23}. These studies have interpreted these discrepancies as indications of a deficiency in the modelling of dust and haze within the synthetic atmosphere models.} 

\textcolor{black}{With the wavelength coverage and data quality of the spectra now provided by JWST, it is imperative to establish a comprehensive understanding of the limitations of self-consistent models of atmospheres and to develop analysis methods that take appropriate account of their systematic errors. This is crucial to ensure the production of reliable and robust estimates of the atmospheric properties of planetary mass objects, and to avoid over-interpretation of these properties, especially when dealing with spectra covering a narrower spectral interval. In this paper, we exploit the unprecedented data provided by JWST to propose an original method that aims to identify and propagate the systematic errors of the models directly into the error bars of the atmospheric parameters estimated from the forward modelling approach. Five different grids of pre-computed synthetic spectra are used, generated from five different self-consistent models, to compare the results they yield and the interpretation that can be made to constrain the formation pathway of VHS~1256~b. Section \ref{sec:data_mod} describes the data set and models used in this study. Section \ref{sec:windows_fit} exploits the broad wavelength coverage provided by JWST to study the dispersion of parameter estimates along the spectral energy distribution (SED). Section \ref{sec:features_fit} focuses on medium-resolution spectral features to estimate chemical abundances. Finally, Section \ref{sec:discussion} is devoted to a discussion of the results, and a conclusion is given in Section \ref{sec:summary}.}
  
\section{Description of the data and the models}
\label{sec:data_mod}

\begin{figure}[!t]
\centering
\includegraphics[width=1.0\hsize]{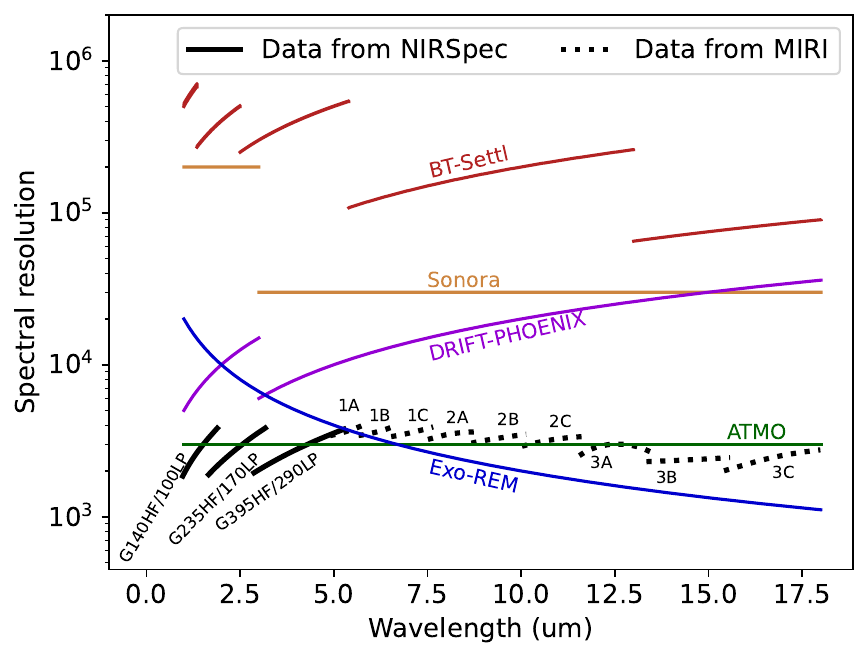}
    \caption{Spectral resolution as a function of the wavelength coverage allowed by the NIRSpec + MIRI data. The observed data are represented by the black line. The five models are depicted by the colored lines: \texttt{ATMO} in green; \texttt{Exo-REM} in blue; \texttt{Sonora} in yellow; \texttt{BT-Settl} in red; and \texttt{DRIFT-PHOENIX} in purple. For the models, the spectral resolution has been calculated assuming a Nyquist sampling rate.}
    \label{fig:resolution}
\end{figure}

\subsection{Data}
\label{sec:data}

\textcolor{black}{The dataset analyzed in this study was first presented in \cite{Miles22} as the first spectroscopic dataset of a substellar object obtained through direct imaging with JWST (ERS \#1386, \citealt{Hinkley22}). The observations were conducted using two instruments: the Near-Infrared Spectrograph (NIRSpec, \citealt{Jakobsen22, Boker22}) in Integral Field Unit (IFU) mode and the Mid-Infrared Instrument (MIRI, \citealt{Wright23}) in Medium Resolution Spectrometer (MRS, \citealt{Argyriou23}) mode. This dataset offers the widest wavelength coverage to date for this kind of object, spanning from $\sim$ 1 to 18~\mic. It is important to note that the signal-to-noise ratio (SNR) dropped significantly in channels 4A, 4B, and 4C of MIRI, which covers wavelengths above 18~\mic. As a result, this channel was not included in this analysis. The spectral resolution of the data ranges from $\sim$ 1500 to 3500 (see Figure \ref{fig:resolution} for a visualization of the resolution). The dataset is composed of twelve individual channels, with three acquired by NIRSpec and nine obtained by MIRI. These channels were sequentially observed over a time span from UT 09:43:42 to 13:56:05 on 2022-07-05, corresponding to $\sim$~20\% of the rotation period of the object. The data were reduced using adapted version 1.7.2 of the standard JWST pipeline for NIRSpec \citep{calib_jwst_172} and adapted version 1.8.1 for MIRI \citep{calib_jwst_181}. The different steps of the reduction are detailed in \cite{Miles22}.}

\subsection{Models}
\label{sec:mod}

\begin{table*}[t!]
\caption{Parameter spaces explored by the models: effective temperature (\Teff); surface gravity (log(g)); metallicity ([M/H]); carbon-oxygen ratio (C/O); adiabatic index ($\gamma$); and sedimentation factor (\fsed).}
\label{tab:param_model}
\renewcommand{\arraystretch}{1.31}
\begin{center}
\small
\begin{tabular}{c||cc|cc|cc|cc|cc|cc}
\hline
\hline
                        & \multicolumn{2}{|c}{\Teff} & \multicolumn{2}{|c}{log(g)} & \multicolumn{2}{|c}{[M/H]} & \multicolumn{2}{|c}{C/O} & \multicolumn{2}{|c}{$\gamma$} & \multicolumn{2}{|c}{\fsed} \\
                        & \multicolumn{2}{|c}{(K)}   & \multicolumn{2}{|c|}{(dex)}  &           &               &             &             &             &                &             &             \\
                        & range         &     step   & range     &     step       & range     &     step      & range       &     step    & range       &     step       & range       &     step    \\
\hline
\texttt{ATMO}           & [800,3000]    & 100        & [2.5,5.5] & 0.5            & [-0.6,0.6]& 0.3           & [0.3,0.7]   & 0.25        & [1.01,1.05] & 0.02           & -           & -           \\
\texttt{Exo-REM}        & [400,2000]    & 100        & [3.0,5.0] & 0.5            & [-0.5,0.5]& 0.5           & [0.1,0.8]   & 0.1         & -           & -              & -           & -           \\
\texttt{Sonora}         & [900,2400]    & 100        & [3.5,5.5] & 0.5            & [-0.5,0.5]& 0.5           & -           & -           & -           & -              & [1.0,8.0]   & 1.0         \\
\texttt{BT-Settl}       & [1200,3000]   & 100        & [2.5,5.5] & 0.5            & -         & -             & -           & -           & -           & -              & -           & -           \\
\texttt{DRIFT-PHOENIX}  & [1000,3000]   & 100        & [3.0,6.0] & 0.3            & [-0.6,0.3]& 0.3           & -           & -           & -           & -              & -           & -           \\
\hline
\hline
\end{tabular}
\end{center}
\end{table*}

\textcolor{black}{Since the initial application of atmospheric models to imaged planetary mass objects in the late 1990s \citep{Allard96, Marley96, Tsuji96}, various families of 1-D models have been developed to replicate the spectral characteristics of progressively colder atmospheres. To address current challenges in atmospheric modelling (cloud formation and evolution, non-equilibrium chemistry, etc.) these models employ different strategies, varying in the complexity and inclusion of physical processes, as well as the range of parameters they explore. Additionally, these models can cover different wavelength ranges and operate at various spectral resolutions. In this study, we considered five distinct grids of synthetic spectra, generated by four cloudy and one cloud-free model. The rest of this section is dedicated to their introduction. The parameter spaces they investigate are summarized in Table \ref{tab:param_model}. 
Their spectral resolutions have been estimated from the sampling of the synthetic spectra they provided, assuming a Nyquist sampling rate. They are illustrated in Figure \ref{fig:resolution}.}

\begin{itemize}
    
    \item \underline{\texttt{ATMO}} \citep{Tremblin15}: \textcolor{black}{This is a cloudless model that focuses on two non-equilibrium chemical reactions: CO/CH$_{4}$ and N$_{2}$/NH$_{3}$. These instabilities can induce diabatic convection (fingering convection), due to the difference in mean-molecular weights between CO and CH$_{4}$ at the L/T transition, and N$_{2}$ and NH$_{3}$ at the T/Y transition. That affects the temperature-pressure (T-P) profile in the atmosphere, reducing the temperature gradient, and providing an alternative explanation for observed reddening that does not invoke clouds. The adiabatic index $\rm \gamma$ drives this temperature gradient, while log(g) determines the eddy coefficient responsible for vertical mixing (and thus non-equilibrium chemistry). This approach is debated \citep{Leconte18} and the lack of clouds is now challenged by the clear detection of silicate features. But the output spectra reproduce very well the observations, in particular at the L/T transition (e.g., \citealt{Petrus23}) and it remains to be understood how far this proposed ingredient could play a role in changing the T-P profile and producing redder NIR and dimmer spectra, as produced by \texttt{ATMO}. This model incorporates 277 species, assuming elemental abundances from \cite{Caffau11}, and including those related to non-equilibrium chemistry as defined by the model of \cite{Venot12} originally developed for hot Jupiters, while the other species precipitate under the photosphere. Opacity sources encompass collision-induced absorptions of H$_{\rm 2}$–H$_{\rm 2}$ and H$_{\rm 2}$–He, thirteen molecules (H$_{\rm 2}$O, CO$_{\rm 2}$, CO, CH$_{\rm 4}$, NH$_{\rm 3}$, TiO, VO, FeH, PH$_{\rm 3}$, H$_{\rm 2}$S, HCN, C$_{\rm 2}$H$_{\rm 2}$, and SO$_{\rm 2}$), seven atoms (Na, K, Li, Rb, Cs, Fe, and H$^{\rm -}$), and the Rayleigh scattering opacities for H$_{\rm 2}$, He, CO, N$_{\rm 2}$, CH$_{\rm 4}$, NH$_{\rm 3}$, H$_{\rm 2}$O, CO$_{\rm 2}$, H$_{\rm 2}$S, SO$_{\rm 2}$. This model explores five parameters: \Teff~from 800 to 3000 K, log(g) from 2.5 to 5.5 dex, [M/H] from -0.6 to 0.6, C/O from 0.3 to 0.7, and $\gamma$ from 1.01 to 1.05.}

    \item \underline{\texttt{Exo-REM}} \citep{Charnay18}: \textcolor{black}{This model adopts a simplified approach to microphysics to facilitate constraining and interpreting atmospheric parameters. It calculates the flux iteratively through 65 sub-layers of the photosphere under the assumption of radiative-convective equilibrium. Non-equilibrium chemistry is considered, following \cite{Zahnle14}, for a limited number of molecules (CO, CH$_{\rm 4}$, CO$_{\rm 2}$, and NH$_{\rm 3}$) and the cloud model incorporates the formation of iron, Na$_{\rm 2}$S, KCl, silicates, and water. Opacity sources include collision-induced absorptions of H$_{\rm 2}$–H$_{\rm 2}$ and H$_{\rm 2}$–He, rovibrational bands from nine molecules (H$_{\rm 2}$O, CH$_{\rm 4}$, CO, CO$_{\rm 2}$, NH$_{\rm 3}$, PH$_{\rm 3}$, TiO, VO, and FeH), and resonant lines from Na and K. Vertical mixing is parameterized using an eddy-mixing coefficient derived from cloud-free simulations. The chemical abundances of each element are defined according to \cite{Lodders10}. By simulating the dominant physical and chemical processes and incorporating the feedback effect of clouds on the thermal structure of the atmosphere, \texttt{Exo-REM} reproduces the spectra of objects at the L-T transition. The parameter space explored includes \Teff~ranging from 400 to 2000~K, log(g) from 3.0 to 5.0~dex, [M/H] from -0.5 to 0.5, and C/O from 0.1 to 0.8.}

    \item \underline{\texttt{Sonora}} (Morley et al., in prep.): \textcolor{black}{\texttt{Sonora Diamondback} is a new grid of models within the \texttt{Sonora} family of models (\texttt{Sonora-Bobcat}: \citealt{Marley21}; \texttt{Sonora-Cholla}: \citealt{Karalidi21}). It is based on the radiative-convective equilibrium model described in \citealt{Marley99}, and used to model brown dwarfs and exoplanets (e.g., \citealt{Saumon08}, \citealt{Fortney08}, \citealt{Morley12}). Clouds are parameterized following the approach in \cite{Ackerman01}. Opacities are included for 15 molecules and atoms, as well as collision-induced opacity of hydrogen and helium and the solar abundances from \cite{Lodders10} are considered. Vertical mixing in the cloud model is calculated using the mixing length theory. Models assume chemical equilibrium throughout the atmosphere. The parameter space considered includes temperatures from 900 to 2400~K, log(g) from 3.5 to 5.5~dex, [M/H] from -0.5 to +0.5, and the cloud sedimentation efficiency parameter \fsed from 1 (full cloudy) to 8. (cloud-less).}

    \item \underline{\texttt{BT-Settl}} \citep{Allard12}: \textcolor{black}{This model estimates the abundance and size distributions of dust grains by comparing timescales of condensation, coalescence, mixing, and gravitational settling for 55 types of solids in different atmospheric layers. The radiative transfer is then calculated using the \texttt{PHOENIX} code. Non-equilibrium chemistry is permitted for several molecules, including CO, CH$_{\rm 4}$, CO$_{\rm 2}$, N$_{\rm 2}$, and NH$_{\rm 3}$. Vertical mixing is accounted for using the mixing-length theory under hydrostatic and chemical equilibrium. We used the \texttt{CIFIST} version of \texttt{BT-Settl}, which incorporates solar chemical abundances defined by \cite{Caffau11}. The explored parameter space includes \Teff~ranging from 1200 to 7000~K (with the high end limited to 3000~K for computational efficiency) and log(g) ranging from 2.5 to 5.5~dex.}

    \item \underline{\texttt{DRIFT-PHOENIX}} \citep{Helling08, Witte09, Witte11}: \textcolor{black}{This model combines the stationary non-equilibrium cloud model called \texttt{DRIFT} \citep{Woitke03, Woitke04, Helling06} that incorporates growth, evaporation, and gravitational settling of seven solids (TiO$_{2}$, Al$_{2}$O$_{3}$, Fe, SiO$_{2}$, MgO, MgSiO$_{3}$, Mg$_{2}$SiO$_{4}$), with the code \texttt{PHOENIX} \citep{Hauschildt97, Allard01} that calculates radiative transfer, hydrostatic and chemical equilibrium, and employs mixing-length theory in 256 atmospheric layers. To replenish the upper layers depleted of grains due to condensation, vertical mixing by convection is included. The grain coagulation is not taken into account. The parameter space covered by this model ranges from \Teff~ of 1000 to 3000~K, log(g) from 3.0 to 6.0~dex, and metallicity [M/H] from -3.0 to 6.0. For [M/H]~=~0.0 the solar element abundances from \cite{Grevesse92} are used, implying a fixed C/O~=~0.55 in the grid.}

\end{itemize}

\subsection{Parameter space exploration with \texttt{ForMoSA}}
\textcolor{black}{To compare these models with the data, we used the forward modelling code \texttt{ForMoSA}\footnote{https://formosa.readthedocs.io}. Initially introduced in \cite{Petrus20}, this code has been updated for the analysis of the VLT/X-Shooter data related to VHS~1256~b \citep{Petrus23}. This updated version is the one considered in this work. This publicly available tool utilizes the Bayesian inversion technique known as "nested sampling" \citep{Skilling04} to efficiently explore the complex parameter space provided by grids of pre-computed synthetic spectra. By doing so, it proposes an estimation of the atmospheric properties of brown dwarfs and directly imaged planets, but also other parameters such as the radius, the interstellar extinction, the radial velocity, and the projected rotational velocity. Regarding the spectral resolution of the data analyzed in this study, only the radius will be constrained here, in addition to the parameters explored by the grids. In order to optimize the analysis process, the model grids presented in this study have been converted to the \texttt{xarray}\footnote{https://docs.xarray.dev} format. This format allows for efficient manipulation, including interpolation and labeling of the various dimensions allowed by the grids. The spectral resolution of both the data and synthetic spectra is compared at each wavelength to align them with optimized values. When the model has a lower spectral resolution compared to the data (as observed in certain \texttt{ATMO} and \texttt{ExoREM} wavelengths, see Figure \ref{fig:resolution}), the data's resolution is adjusted accordingly. This adjustment involves convolving both the observed and synthetic spectra with a Gaussian distribution, using the calculated FWHM to achieve the desired resolution. Subsequently, the \texttt{Python} module \texttt{spectres}\footnote{https://pypi.org/project/spectres/} is employed to resample the spectra onto a wavelength grid, ensuring a defined Nyquist sampling and achieving the desired spectral resolution with wavelengths. This final step optimizes computation time by limiting the number of data points considered during Bayesian inversion, all while preserving the information. The nested sampling inversion process is then carried out using the \texttt{Python} module \texttt{nestle}\footnote{http://kylebarbary.com/nestle/}. For each fit, we considered priors uniformly distributed through the parameter space defined by the grids, a uniform prior between 0 and 10~\Rjup for the radius, and a Gaussian prior for the distance with $\rm \mu$=21.15~pc and $\rm \sigma$=0.22~pc, corresponding to the Gaia DR3 measurement and error, respectively.}

\section{Fits as a function of the wavelength range}
\label{sec:windows_fit}

\begin{figure*}[hbtp!]
\vspace{0.5cm}
\centering
\includegraphics[width=1.0\hsize]{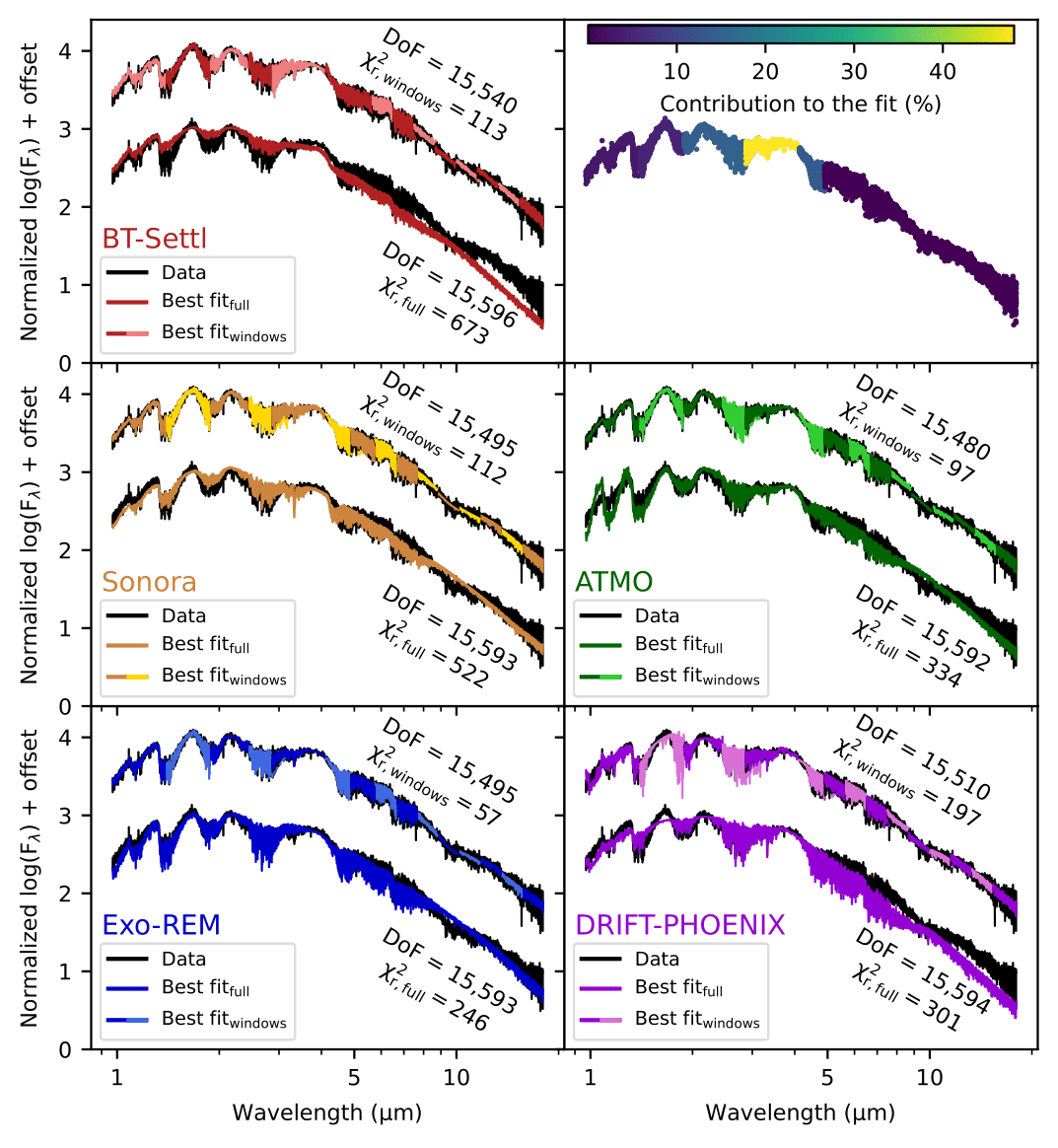}
    \caption{Comparison between the data and the interpolated synthetic spectra generated with the set of parameters that maximized the likelihood for each model. The best fit using the full SED is depicted with unicolor lines, while the best fits using different spectral windows are illustrated with bi-color lines. The data is represented by black lines. All spectra have been normalized with the bolometric luminosity of VHS~1256~b to simplify the figure, and an offset was applied. The use of spectral windows enhances the quality of the fit for each model. We note that a possible underestimation of the error bars of the data may play a role in the high value of \chr, which is provided as a comparative measure of the performance of the best-fit models, as well as the number of degrees of freedom (DoF) that have been used to calculate it. The top-right panel shows the relative contribution of each spectral window to the fit using the full SED. It corresponds to $C_{win}$/$C_{win_tot}$ when $C_{win}$ is calculated from Equation \ref{eq:cwin} and $C_{win_tot}$ is the sum of $C_{win}$ over the 15 spectral windows. The NIRSpec channel G395HF/290LP, from 2.87 to 5.27~\mic, with its higher spectral resolution and signal-to-noise ratio, carries the greatest weight in the calculation of the likelihood. See Appendix \ref{App:zoom} for a zoomed version of this plot.}
    \label{fig:fits}
    
\end{figure*}

\textcolor{black}{The forward modelling approach used in this study offers several advantages, including the ability to fit high-resolution data across a broad range of wavelengths within a relatively short computing time (less than 1 day). However, this approach has its challenges, as the quality of the fit is directly influenced by the quality of the model, which can be difficult to fully quantify. \cite{Petrus20} demonstrated that systematic errors present in the model \texttt{BT-Settl} greatly influenced the posteriors of the parameter estimates explored during the inversion of 9 spectra of young M7-M9 dwarfs. The fits converged to different sets of parameters depending on the spectral band considered. These differences considerably dominated the errors on these parameters derived from the Bayesian algorithm. This work further extended to the analysis of isolated brown dwarfs at the L/T transition observed with Spitzer \citep{Suarez21}, L dwarfs using VLT/X-Shooter data of VHS~1256~b with the \texttt{ATMO} grid \citep{Petrus23}, and VLT/SINFONI data of AB~Pic~b with the \texttt{BT-Settl} and \texttt{Exo-REM} grids \citep{Palma23}. These subsequent studies confirmed that the systematic errors observed in the previous work were indeed a recurring issue in the forward modelling approach. In this section, we present an extension of the method proposed in these three previous studies. We address the systematics in the models to provide a robust estimation of the atmospheric properties of VHS~1256~b, taking advantage of the extensive spectral coverage offered by JWST.}

\subsection{Description of the method}
\label{sec:windows_fit_meth}

\textcolor{black}{To assess the influence of systematic effects on the estimation of parameters, we conducted a comparative analysis between the results obtained from fitting the full wavelength coverage permitted by JWST and the results obtained from fitting different spectral windows defined along the SED. We applied each model introduced in Section \ref{sec:mod} for this purpose.}

\textcolor{black}{In the initial step, we combined the NIRSpec and MIRI spectra. For regions where the wavelength coverage overlapped between the channels, we selected the data with the highest spectral resolution. The fitting procedure was then carried out for each model, considering a wavelength range from 0.97 to 18.02 \mic. The corresponding fits are represented by uni-color solid lines in Figure \ref{fig:fits} and the corresponding sets of parameters are summarized in Tables \ref{tab:final_part1_grid} and \ref{tab:final_part1_extragrid}.}

\textcolor{black}{Next, we divided the spectrum into 15 distinct spectral windows and conducted independent fits for each window. To account for potential calibration discrepancies between channels, we utilized the wavelength coverage of each channel to define the specific range for each window. For the NIRSpec data, we further divided them into sub-windows to examine the model performance across different infrared bands, while in the case of MIRI, each channel was treated as an individual spectral window. The complete list of spectral windows can be found in Tables \ref{tab:spec_wind_nirspec} and \ref{tab:spec_wind_miri}, and the corresponding best fits and parameters estimates are illustrated in Figure \ref{fig:fits} using two-color lines and Tables \ref{tab:final_part1_grid} and \ref{tab:final_part1_extragrid}, respectively.}

\begin{table}[h]
\caption{Spectral windows defined for the NIRSpec data. Each NIRSpec channel has been split into two windows, corresponding to different NIR bands.}
\label{tab:spec_wind_nirspec}
\renewcommand{\arraystretch}{1.31}
\begin{center}
\small
\begin{tabular}{cc|cc}
\hline
\hline
\multirow{2}{*}{NIRSpec}    &        & window 1 & window 2   \\
&                     & (\mic) & (\mic)   \\
\hline
\multirow{3}{*}{\rotatebox[origin=c]{90}{Channels}} & G140HF/100LP  & 0.97-1.40 & 1.40-1.89  \\
 & G235HF/170LP  & 1.89-2.45 & 2.45-3.17   \\
 & G395HF/290LP  & 2.87-4.10 & 4.10-5.27   \\
\hline
\hline
\end{tabular}
\end{center}
\end{table}

\begin{table}[h]
\caption{Spectral windows defined for MIRI data. The nine windows correspond to the three channels used to acquire the data combined with the three grating settings.}
\label{tab:spec_wind_miri}
\renewcommand{\arraystretch}{1.31}
\begin{center}
\small
\begin{tabular}{cc|ccc}
\hline
\hline
&                  & \multicolumn{3}{|c}{Grating settings} \\
MIRI  &            & A & B & C  \\
&                  & (\mic) & (\mic) & (\mic)  \\
\hline
\multirow{3}{*}{\rotatebox[origin=c]{90}{Channels}} & 1 & 4.90-5.74 & 5.66-6.63 & 6.53-7.65  \\
                                                   & 2 & 7.51-8.77 & 8.67-10.13 & 10.02-11.70  \\
                                                   & 3 & 11.55-13.47 & 13.34-15.57 & 15.41-17.98\\
\hline
\hline
\end{tabular}
\end{center}
\end{table}

\textcolor{black}{One of the advantages of this approach is the ability to assess the contribution of each spectral window to the likelihood calculation, denoted as $C_{\text{win}}$. This allows us to identify the spectral windows that have the most significant impact on the inversion results when considering the full SED. The calculation of $C_{\text{win}}$ is performed using the following formula:}

\begin{equation}
    \label{eq:cwin}
    C_{win} \propto \sum_{\lambda} SNR_{\lambda}^{2}
\end{equation}

\textcolor{black}{where $SNR_{\lambda}$ is the signal-to-noise ratio associated with each wavelength, assuming completely uncorrelated errors. The sum of $SNR^{2}$ is considered to align with the definition of the likelihood function utilized during the fits, which is directly proportional to a $\chi^2$ value. Consequently, $C_{win}$ is maximized for spectral windows that have a greater number of data points and higher SNR. This information is illustrated in the top-right panel of Figure \ref{fig:fits}. In the case of JWST's data, the fit is primarily driven by the NIRSpec G395HF/290LP channel, which contributes approximately 60\% to the overall fit. On the other hand, all MIRI data collectively contribute around 0.26\% to the overall fit.}

\textcolor{black}{In this approach, we also gain the ability to estimate the sensitivity of the models to each parameter depending on the spectral window used. For a given spectral window $\Delta\uplambda$, model $m$, and parameter $p$, we calculated this sensitivity $S_{\Delta\uplambda,m,p}$ using the following relation:}

\begin{equation}
\label{eq:sensibility}
    S_{\Delta\uplambda,m,p} = \sum_{\uplambda} \left( \frac{abs(F_{m, \uplambda}(p_{max})-F_{m, \uplambda}(p_{min}))}{F_{m, \uplambda}(p_{max})} \right) \times \frac{1}{N_{\Delta\uplambda}}
\end{equation}

\textcolor{black}{where $F_{m, \uplambda}(p_{min})$ and $F_{m, \uplambda}(p_{max})$ are the synthetic flux generated by the model $m$, at the wavelength $\uplambda$, with the minimum and maximum values of the parameter $p$ explored by $m$, respectively, and $N_{\Delta\uplambda}$ is the number of data points contained in $\Delta\uplambda$. The remaining parameters are estimated using the posteriors of the fits corresponding to the specific spectral window and model considered. Therefore, these index values are only relevant for objects with atmospheric properties similar to VHS~1256~b. If, for a given model $m$, the spectral information $F_{m, \uplambda}(p)$ within the considered window is not affected significantly by the parameter $p$, the difference $F_{m, \uplambda}(p_{max})$ - $F_{m, \uplambda}(p_{min})$ (and consequently $S_{\Delta\uplambda,m,p}$) will tend toward 0. Conversely, a substantial $S_{\Delta\uplambda,m,p}$ value will result if there's significant spectral variation within the window corresponding to variations in the parameter $p$. The sensitivity is visualized in Figures \ref{fig:teff_logg_compare}-\ref{fig:gammafsed_compare} through the size of the data points, and will be further discussed in the subsequent sections.}


\subsection{Results: Parameters as a function of wavelengths}
\label{sec:res_part1_final}
\textcolor{black}{All five models exhibit deviations from the observed data when the entire wavelength range is considered for the fit, particularly at longer wavelengths ($\uplambda >$ 7 \mic), which carry less weight in the likelihood calculation. To assess the goodness of fit, we calculated the reduced $\chi^2$ values for the fits using the full SED, denoted as $\chi_{r, \text{full}}^2$, as well as for the fits using the spectral windows, denoted as $\chi_{r, \text{win}}^2$ (see Figure \ref{fig:fits}). For the latter case, we constructed a synthetic SED by combining the best fit from each spectral window, following the merging procedure described earlier, enabling a direct comparison with the fit of the full SED. Among the models tested, the \texttt{Exo-REM} model demonstrates the best performance, yielding $\chi_{r, \text{full}}^2$~=~246 and $\chi_{r, \text{win}}^2$~=~57. The utilization of smaller wavelength ranges in the independent fits (spectral windows) results in a decrease in the overall $\chi_{r, \text{win}}^2$ from a factor of 1.5 for the \texttt{DRIFT-PHOENIX} model to 6 for the \texttt{BT-Settl} model.  This improvement is not surprising, as this procedure is similar to a fit in which the number of free parameters is multiplied by the number of spectral windows. Consequently, there is a variation in the estimated parameters depending on the specific spectral window considered.} 

\begin{figure*}[!t]
\centering
\includegraphics[width=0.49\hsize]{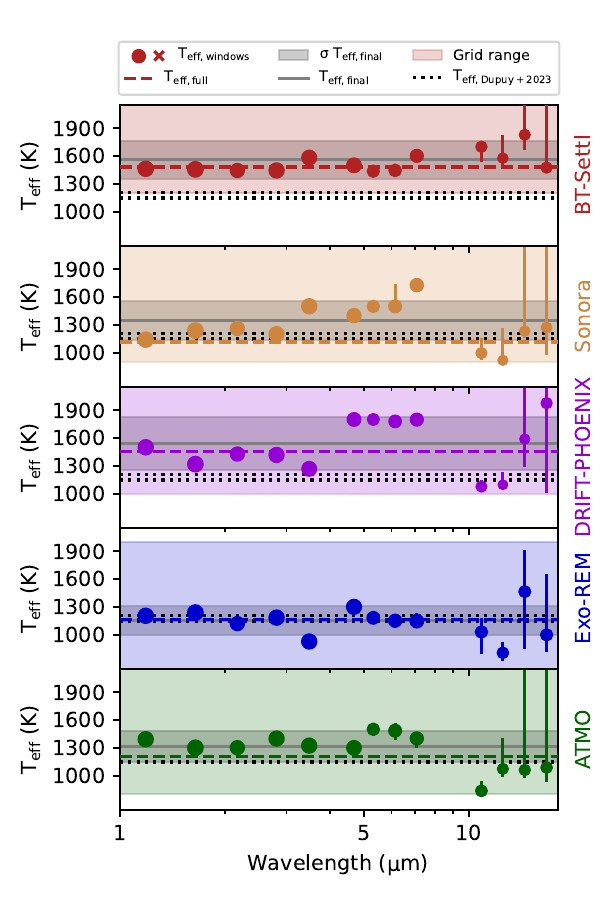}
\includegraphics[width=0.49\hsize]{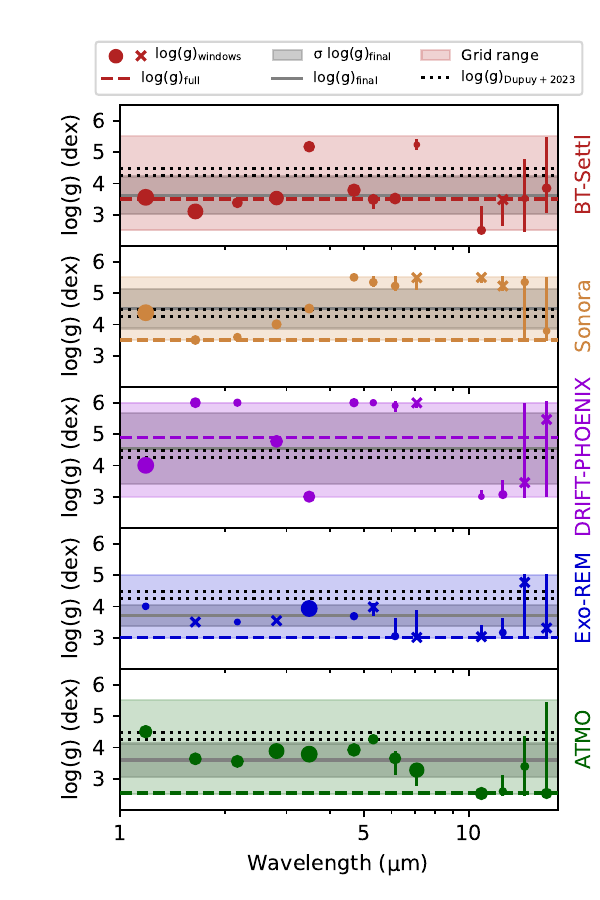}
    \caption{Estimation of the \Teff (left panel) and log(g) (right panel) using the full SED during the fit (dashed colored line) and using the spectral windows (colored dots). The size of the dots indicates the sensitivity of each spectral window to each parameter. If the sensitivity is too low to be represented by a dot, a colored cross is used. Each panel corresponds to a different model, with the parameter range indicated by the filled colored area. The final value and error extracted from these results are shown by the solid gray line and gray area, respectively. The dotted black lines represent the value estimated by \cite{Dupuy23}, who interpolated the evolutionary model from \cite{Saumon08} assuming the bolometric luminosity and age of VHS~1256~b. The two solutions they have found are given.}
    \label{fig:teff_logg_compare}
\end{figure*}

\textcolor{black}{As illustrated in Figures \ref{fig:teff_logg_compare} to \ref{fig:gammafsed_compare}, the high signal-to-noise ratio provided by JWST allows for precise estimation of the atmospheric properties of VHS~1256~b in each fit, resulting in very small error bars. However, the significant dispersion observed in these estimates across the different spectral windows indicates that these errors cannot be representative of this dispersion. Furthermore, as depicted in these figures, the sensitivity of the models to each parameter varies depending on the spectral window considered. We propose here a procedure that provides a robust estimate of each parameter $\rm \Theta_{win,f}$, where $\rm \Theta$ represents the parameter of interest. This procedure considers both the dispersion and the models' sensitivity of each parameter to redefine a more robust error bar.}

\textcolor{black}{For each parameter estimated with each model, we defined a sample composed of subsamples of points randomly drawn from the posteriors obtained with each spectral window. We excluded those affected by the silicate absorption between 7.5 and 10.5~\mic which are known to be not reproduced by the current generation of models. The number of points in each subsample is determined proportionally based on the calculated sensitivity $S_{\Delta\uplambda,m,p}$ using Equation \ref{eq:sensibility}. The final value and error of each parameter are defined as the mean and standard deviation, respectively, of this sample, and are presented in Figures \ref{fig:teff_logg_compare}-\ref{fig:gammafsed_compare} as the solid gray line and gray area, respectively. They are also reported in Tables \ref{tab:final_part1_grid} and \ref{tab:final_part1_extragrid}.}


\subsubsection{The effective temperature \Teff}
\textcolor{black}{Considering the young age of VHS~1256~b, the estimate of the T$_{\rm eff, full}$ considering the full wavelength range, is consistent with its spectral type and with the prediction of the evolutionary models \citep{Dupuy23} for the grids \texttt{Sonora}, \texttt{Exo-REM}, and \texttt{ATMO} (Figure \ref{fig:teff_logg_compare}). The \texttt{BT-Settl} and \texttt{DRIFT-PHOENIX} grids appear to converge towards higher \Teff values than those predicted by evolutionary models. However, it is noteworthy that for these two models, \texttt{ForMoSA} converges to \Teff values close to their lower limit. The \texttt{ATMO}, and \texttt{Exo-REM} grids demonstrate greater stability across the different spectral windows compared to \texttt{BT-Settl}, \texttt{Sonora}, and \texttt{DRIFT-PHOENIX}. This leads to smaller error bars on the final values T$_{\rm eff, win, f}$ which remains consistent with T$_{\rm eff, full}$ for each model. The large error bar at $\rm \lambda >$13~\mic is attributed to the low signal-to-noise ratio at these wavelengths. The sensitivity of \Teff remains relatively constant across all spectral windows.}

\subsubsection{The surface gravity log(g)}

\textcolor{black}{The results using the full wavelength range for the models \texttt{Exo-REM}, \texttt{ATMO}, \texttt{Sonora}, and \texttt{BT-Settl} converge towards low values of log(g) ($<$4.5~dex), whereas \texttt{DRIFT-PHOENIX} tends to converge towards high values ($>$4.5~dex; Figure \ref{fig:teff_logg_compare}). Significant dispersion was observed across the various spectral windows, covering the entire range of log(g) values explored, particularly for the \texttt{BT-Settl}, \texttt{Sonora}, and \texttt{DRIFT-PHOENIX} grids. Smaller dispersion was also evident in the other grids \texttt{Exo-REM} and \texttt{ATMO}. However, trends could be identified, with a preference for lower surface gravities along the SED. The grid \texttt{Sonora} provides low log(g) for wavelengths shorter than 4~\mic but higher log(g) for longer wavelengths. The sensitivity analysis confirmed that the log(g) parameter is primarily constrained by the near-infrared (NIR) data ($\uplambda < $ 5~\mic). Due to the wide dispersion observed across the different spectral windows, it is challenging to provide a robust value for log(g) except for \texttt{Exo-REM} and \texttt{ATMO}, which converge to consistent values. Although these log(g)$_{\rm win, f}$ point to low values ($<$ 4.12~dex), typical of young objects, they are significantly lower than the predictions of evolutionary models provided by \cite{Dupuy23}.}


\begin{figure*}[!t]
\centering
\includegraphics[width=0.49\hsize]{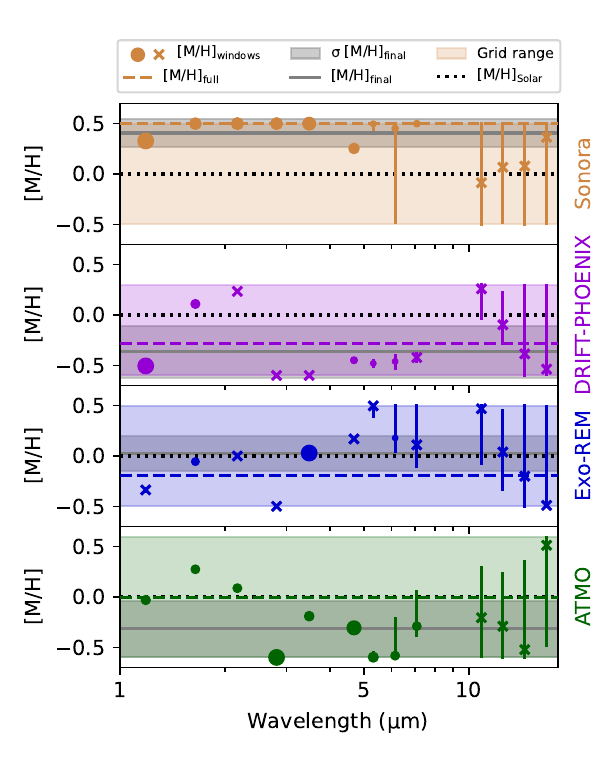}
\includegraphics[width=0.49\hsize]{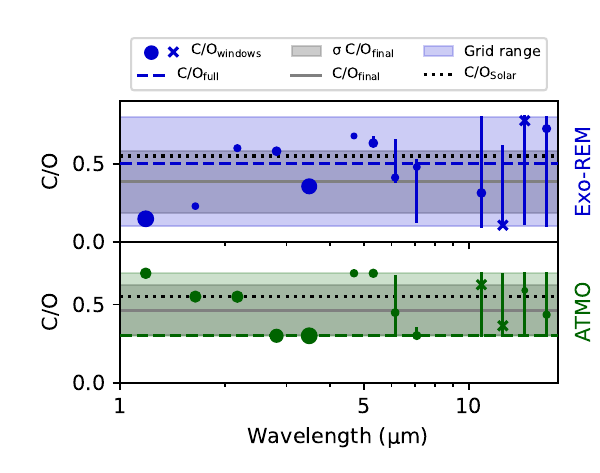}
    \caption{Same figure as Figure \ref{fig:teff_logg_compare} but for the [M/H] (left) and the C/O (right). The dotted black lines represent the solar values 0.0 and 0.55 for [M/H] and C/O, respectively.}
    \label{fig:mhco_compare}
\end{figure*}

\subsubsection{The metallicity [M/H]}
\textcolor{black}{The exploration of [M/H] is performed using the grids \texttt{Exo-REM}, \texttt{Sonora}, \texttt{DRIFT-PHOENIX}, and \texttt{ATMO}. When the full wavelength range is considered, the \texttt{ATMO} model indicates a solar [M/H], while the \texttt{Sonora} model suggests a supersolar metallicity. On the other hand, both the \texttt{Exo-REM} and \texttt{DRIFT-PHOENIX} models converge towards a subsolar metallicity (Figure \ref{fig:mhco_compare}). These results vary when considering the spectral windows. In this case, the \texttt{Exo-REM} model estimates a solar [M/H], while the \texttt{Sonora} model suggests a supersolar [M/H]. Conversely, both the \texttt{DRIFT-PHOENIX} and \texttt{ATMO} models converge towards a subsolar [M/H]. The three identified regimes of metallicity are further confirmed by the estimates of [M/H]$_{\rm win,f}$. Similar to the log(g) parameter, the [M/H] estimation is primarily constrained by the near-infrared (NIR) data ($\uplambda < $ 5~\mic).}

\subsubsection{The carbon-oxygen ratio C/O}
\textcolor{black}{Among the considered grids, only \texttt{Exo-REM} and \texttt{ATMO} enable the exploration of the C/O ratio (Figure \ref{fig:mhco_compare}). The \texttt{Exo-REM} model appears to exhibit a preference for low C/O ratios, although the wide dispersion of estimates makes it difficult to establish a definitive trend. The situation is even more complex with the \texttt{ATMO} model, as the estimated C/O ratios span the entire range of values explored by the grid. Consequently, this parameter cannot be reliably estimated using this method. Nonetheless, it appears that the NIR data exhibit the highest sensitivity to this parameter} 

\begin{figure*}[!t]
\centering
\includegraphics[width=0.49\hsize]{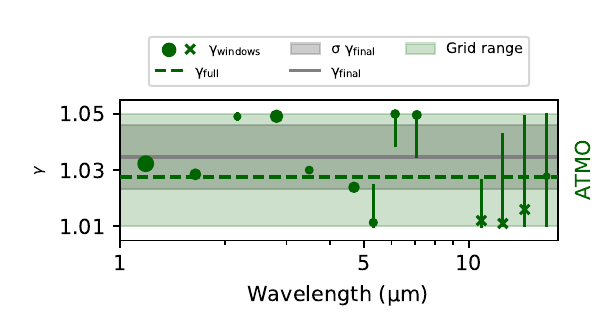}
\includegraphics[width=0.49\hsize]{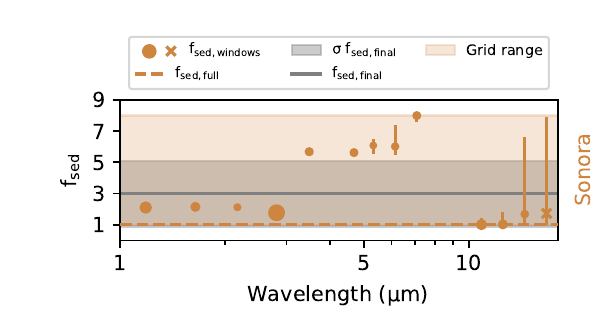}
    \caption{Same figure as Figure \ref{fig:teff_logg_compare} but for the adiabatic index $\rm \gamma$ (left) and the sedimentation factor f$\rm _{sed}$ (right).}
    \label{fig:gammafsed_compare}
\end{figure*}

\subsubsection{The adiabatic index $\rm \gamma$ and the sedimentation factor \fsed}

\textcolor{black}{The parameters $\rm \gamma$ and \fsed are specific to the \texttt{ATMO} and \texttt{Sonora} models, respectively (see Figure \ref{fig:gammafsed_compare}). When considering the NIRSpec data, their estimates exhibit distinct trends, with a medium/high value for $\rm \gamma$ and a low value for \fsed. However, these parameters display similar variations when analyzed with the MIRI data. Specifically, there is an initial increase in both parameters up to $\uplambda \sim$ 9~\mic, followed by a sharp decline in the MIR range. Similarly to C/O, the new errors for $\rm \gamma_{win,f}$, and f$_{\rm sed, win,f}$ both fail to represent the dispersion adequately. The \fsed values are particularly intriguing, indicating a minimal impact of clouds at $\uplambda <$ 3~\mic (\fsed $\sim$ 2), a significant impact at $\uplambda =$ [3-10]\mic (\fsed $>$ 5), and no discernible impact at $\uplambda >$ 10\mic (\fsed $\sim$ 1). This may reveal the complexity of the cloud clover in the atmosphere of VHS~1256~b which stands at the L/T transition. Additionally, the estimation of $\gamma$ ($\sim$ 1.03) and \fsed ($\sim$ 1) using the full SED aligns coherently with the results observed for late-L objects \citep{Stephens09, Tremblin17}.}

\begin{figure}[!t]
\centering
\includegraphics[width=0.98\hsize]{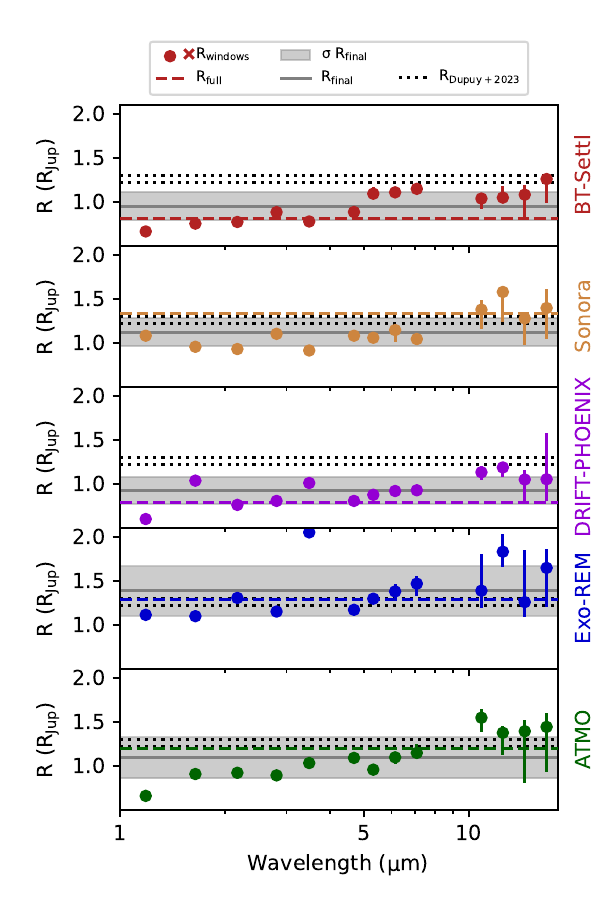}
    \caption{Same figure as Figure \ref{fig:teff_logg_compare}, but for the radius R. This parameter is estimated from the dilution factor $\rm C_{K}$ considering the distance from \citep{Gaia21}.}
    \label{fig:radius_compare}
\end{figure}

\subsubsection{The radius R}
\textcolor{black}{The radius is not a parameter explored by the model grids. Instead, it is estimated using the dilution factor $C_{K} = (R/d)^{2}$, which takes into account the flux dilution between the synthetic spectra generated at the outer boundary of the atmosphere and the observed spectrum. Here, $R$ represents the planet's radius and $d$ represents the distance to the planet. To propagate the error from the distance to the radius, we define a Gaussian prior for the distance centered at 21.14~pc with a standard deviation of 0.22~pc, according to the estimate from \citep{Gaia21}. We impose a flat prior to the radius to leave it unconstrained. The results are illustrated in Figure \ref{fig:radius_compare}. It is observed that the radius estimated by each model is consistently underestimated, except for the \texttt{Exo-REM} grid, which provides a radius in good agreement with the predictions of evolutionary models. When the full SED is used for the fit, the radius estimates from the \texttt{Sonora} and \texttt{ATMO} grids also align with the results from \cite{Dupuy23} and \cite{Miles22}. However, a significant dispersion remains, resulting in substantial error bars for the final values (see Table \ref{tab:final_part1_extragrid}).}


\begin{table*}[h]
\caption{Summary of the estimations of the parameters explored by the grids which are provided for both the full wavelength range and the combined spectral windows. For the combined spectral windows, error bars have been calculated by considering the dispersion of the individual fits from each spectral window. The error bars of the fits on the full SED reflect the high SNR of the data.}
\label{tab:final_part1_grid}
\renewcommand{\arraystretch}{1.31}
\begin{center}
\small
\begin{tabular}{c|c|cccccc}
\hline
\hline
                                        & Fit     & \Teff            & log(g)          & [M/H]               & C/O             & $\gamma$          & \fsed                 \\
                                        &         & (K)              & (dex)           &                     &                 &                   &                       \\
\hline
\multirow{2}{*}{\texttt{ATMO}}           & full    & 1203 $\pm$ 1   & 2.54 $\pm$ 0.01 & -0.01 $\pm$ 0.01    & $<$ 0.30 & 1.030 $\pm$ 0.001 & -                     \\
                                        & windows & 1318 $\pm$ 166   & 3.59 $\pm$ 0.53 & -0.32 $\pm$ 0.28    & 0.46 $\pm$ 0.16 & 1.035 $\pm$ 0.011 & -                     \\
\multirow{2}{*}{\texttt{Exo-REM}}        & full    & 1164 $\pm$ 1   & $<$ 3 & -0.19 $\pm$ 0.01     & 0.50 $\pm$ 0.01 & -                 & -                     \\
                                        & windows & 1153 $\pm$ 152   & 3.70 $\pm$ 0.34 & 0.03 $\pm$ 0.17     & 0.38 $\pm$ 0.20 & -                 & -                     \\
\multirow{2}{*}{\texttt{Sonora}}         & full    & 1116 $\pm$ 1   & $<$ 3.50 & $>$ 0.50     & -               & -                 & 1.01 $\pm$ 0.01        \\
                                        & windows & 1349 $\pm$ 208   & 4.50 $\pm$ 0.63 & 0.41 $\pm$ 0.14     & -               & -                 & 2.97 $\pm$ 2.09        \\
\multirow{2}{*}{\texttt{BT-Settl}}       & full    & 1482  $\pm$ 1  & 3.50 $\pm$ 0.01 & -                   & -               & -                 & -                     \\
                                        & windows & 1560  $\pm$ 203  & 3.62 $\pm$ 0.60 & -                   & -               & -                 & -                     \\
\multirow{2}{*}{\texttt{DRIFT-PHOENIX}}  & full    & 1451 $\pm$ 1   & 4.89 $\pm$ 0.01 & -0.28 $\pm$ 0.01    & -               & -                 & -                     \\
                                        & windows & 1540 $\pm$ 286   & 4.54 $\pm$ 1.13 & -0.36 $\pm$ 0.26    & -               & -                 & -                     \\
\hline
\hline
\end{tabular}
\end{center}
\end{table*}

\begin{table*}[h]
\caption{Summary of the estimations of the parameters not explored by the grids which are provided for both the full wavelength range and the combined spectral windows. The radius R is determined from the dilution factor $\rm C_{K}$, the luminosity L is calculated using the Stefan-Boltzmann law and the estimates of the radius and the \Teff, and the mass M is calculated using the gravitational law and the estimates of the radius and the log(g).}
\label{tab:final_part1_extragrid}
\renewcommand{\arraystretch}{1.31}
\begin{center}
\small
\begin{tabular}{c|c|ccc}
\hline
\hline
                        & Fit            &  R                &  log(L$\rm _{bol}$/L$_{\odot}$)  &  M   \\
                        &                &  (\Rjup)          &                                  &  (M$\rm_{Jup}$)  \\
\hline
\multirow{2}{*}{\texttt{ATMO}}           & full    &  1.19 $\pm$ 0.01  &  -4.57 $\pm$ 0.01  &  0.19 $\pm$ 0.01   \\
                                         & windows &  1.09 $\pm$ 0.23  &  -4.56 $\pm$ 0.20  &  2.24 $\pm$ 1.96  \\
\multirow{2}{*}{\texttt{Exo-REM}}        & full    &  1.29 $\pm$ 0.01  &  -4.56 $\pm$ 0.01  &  0.64 $\pm$ 0.01   \\
                                         & windows &  1.39 $\pm$ 0.28  &  -4.55 $\pm$ 0.14  &  7.87 $\pm$ 6.33  \\
\multirow{2}{*}{\texttt{Sonora}}         & full    &  1.34 $\pm$ 0.01  &  -4.59 $\pm$ 0.01  &  2.17 $\pm$ 0.08    \\
                                         & windows &  1.12 $\pm$ 0.16  &  -4.48 $\pm$ 0.25  &  $<$ 94.40    \\
\multirow{2}{*}{\texttt{BT-Settl}}       & full    &  0.81 $\pm$ 0.01  &  -4.54 $\pm$ 0.01  &  0.80 $\pm$ 0.01  \\
                                         & windows &  0.95 $\pm$ 0.16  &  -4.37 $\pm$ 0.29  &  $<$ 20.90 \\
\multirow{2}{*}{\texttt{DRIFT-PHOENIX}}  & full    &  0.79 $\pm$ 0.01  &  -4.60 $\pm$ 0.01  &  18.73 $\pm$ 0.69   \\
                                         & windows &  0.93 $\pm$ 0.15  &  -4.43 $\pm$ 0.31  &  $<$ 249.42  \\
\hline
\hline
\end{tabular}
\end{center}
\end{table*}

\section{Focusing on spectral features}
\label{sec:features_fit}

\textcolor{black}{The medium spectral resolution provided by JWST has enabled the re-detection of various atomic and molecular absorptions in the spectrum of VHS~1256~b. Notably, the two potassium (KI) doublets at 1.173 and 1.248~\mic are present, previously identified as robust indicators of surface gravity and metallicity through empirical and synthetic analysis of spectra from planetary-mass objects (\citealt{Allers13}, \citealt{Petrus23}). Additionally, we identify carbon monoxide (CO) overtones at approximately 2.3~\mic, which can be used to measure the C/O ratio (\citealt{Konopacky13}, \citealt{GRAVITY20}, \citealt{Petrus21}, \citealt{Hoch22}, \citealt{Petrus23}). The methane (CH$_{4}$) absorption between 2.8 and 3.8~\mic is also detected, confirming the lower resolution detection of these species in the atmosphere of VHS~1256~b by \cite{Miles18} who suggested non-equilibrium chemistry between CO and CH$_{4}$ and a significant vertical mixing (K$_{zz} > $ 10$^{8}$ cm$^{2}$ s$^{-1}$) in the atmosphere to explain the presence of this feature. Finally, this JWST spectrum enabled the detection of a highly resolved CO absorption forest centered at 4.6~\mic, already detected in the low-resolution spectra of isolated brown dwarfs by \cite{Sorahana12}. In this section, we conduct a comprehensive analysis of these prominent spectral features to determine the extent to which they can be utilized to constrain atmospheric properties through the forward modelling approach. We assess the performance of each model described in Section \ref{sec:mod}.}

\subsection{Description of the method}

\textcolor{black}{In one of our previous works, \cite{Petrus21} demonstrated the challenges associated with fitting spectral features detected at medium resolution using pre-computed grids of synthetic spectra. They highlighted that even small inhomogeneities in the grids could remain and propagate through interpolation and introduce biases in the resulting posteriors. Reproducing the depth of the absorption features and accounting for systematics in the continuum shape were also identified as potential sources of bias. To mitigate these issues, we have developed a two-step procedure that analyzes each spectral feature independently.}

\textcolor{black}{We initially conducted a fit on wavelength ranges, $\rm \Delta\uplambda_{continuum}$, that avoided the considered spectral features (red spectral windows in Figure \ref{fig:high_res_wind}). From this fit, we extracted the \Teff, which was then used to define a local pseudo-continuum at the positions of the spectral features. This approach serves as an alternative to the definition of the pseudo-continuum as a low-resolution version of the spectrum (e.g. \citealt{Petrus23}), which may be suitable for higher spectral resolution data but could be unreliable for the resolution of JWST (\Rlam $<$ 3000). Additionally, we extracted the dilution factor $C_{K}$ to ensure that the depth of the absorption is solely related to the atmospheric properties explored by the grid. In the second fit, we only considered the wavelength ranges, $\rm \Delta\uplambda_{fit}$, which focused on the spectral features (green spectral windows in Figure \ref{fig:high_res_wind}). We fixed the values of \Teff and $C_{K}$ to the estimates obtained from the first fit. All spectral windows considered are summarized in Table \ref{tab:features_windows}.}

\begin{table*}[t!]
\caption{Spectral windows were employed to fit the four spectral features: K~I lines, CO overtones, CH$\rm_{4}$ absorption, and CO forest. The $\rm \Delta \uplambda_{continuum}$ is used to estimate the \Teff and the radius. These derived values serve as constraints within the $\rm \Delta \uplambda_{fit}$ for fitting these features.}
\label{tab:features_windows}
\renewcommand{\arraystretch}{1.31}
\begin{center}
\small
\begin{tabular}{c|ccc}
\hline
\hline
Chemical element & $\rm \Delta \uplambda_{continuum}$                 &  $\rm \Delta \uplambda_{fit}$    \\
                 & (\mic)                                           &  (\mic)                           \\
\hline
K~I lines             & [1.160, 1.166] + [1.180, 1.241] + [1.255, 1.260] &  [1.166, 1.180] + [1.241, 1.255]  \\
CO overtones              & [2.250, 2.290] + [2.305, 2.319] + [2.330, 2.400] &  [2.290, 2.305] + [2.319, 2.330]  \\
CH$\rm_{4}$ absorption       & [3.100, 3.300] + [3.375, 3.600]                  &  [3.300, 3.375]                   \\
CO forest              & [4.300, 5.000]                                   &  [4.300, 5.000]                   \\
\hline
\hline
\end{tabular}
\end{center}
\end{table*}

\subsection{Results}

\textcolor{black}{For each spectral feature, we illustrate in Figure \ref{fig:high_res_wind} the synthetic spectrum interpolated from the model grids that maximizes the likelihood when compared to the observed data. The reduced \chr, calculated using $\rm \Delta\uplambda_{fit}$, is provided to quantify and compare the quality of each fit. Its value greater than 1 suggests that the models may have limitations in replicating these spectral signatures and/or that there may have been an underestimation of the error bars during data extraction. The corresponding set of parameters is represented in Figure \ref{fig:high_res}. Similar to the method described in Section \ref{sec:windows_fit}, we have observed variations in the estimates of atmospheric parameters depending on the spectral feature considered for the fit, as well as the model used. Globally, the models are capable of reproducing this information at medium resolution. However, certain spectral features appear to be more challenging to reproduce, such as the potassium line at 1.243~\mic. Now, let us delve into the detailed performance of each individual model.}

\subsubsection{\texttt{ATMO}}
\textcolor{black}{The \Teff derived from the various spectral windows excluding the spectral features, is in agreement with the estimate obtained in Section \ref{sec:windows_fit}, ranging between 1242 and 1520~K. This consistency is expected, as we employed a similar methodology by utilizing a fit based on $\rm \Delta\uplambda_{continuum}$, which encompasses a broad range of wavelengths. The low values of log(g) (less than 4.37~dex) also align well with the age of the system. However, it is worth noting that a remarkably low and unphysical estimate (2.56~dex) is obtained when considering methane absorption. Subsolar metallicities are estimated for the companion, except for the potassium absorption lines, which yielded a supersolar value. The reported C/O ratios range from solar to supersolar, except for the CH$_{4}$ feature that provides a sub-solar value. Furthermore, the adiabatic index ($\gamma$) shows variability across the entire grid size in the different fits, making it challenging to identify a clear trend. Among the models used, \texttt{ATMO} demonstrates the ability to reproduce this information at medium resolution with the lowest \chr. However, it seems to inaccurately reproduce the CO overtone at 2.30~\mic.}

\subsubsection{\texttt{Exo-REM}}
\textcolor{black}{Similarly to \texttt{ATMO}, \texttt{Exo-REM} allowed us to estimate consistent \Teff values, ranging from 1233 to 1460~K, and low log(g) values, lower to 4.17~dex. We also encountered discrepancies with the very low \Teff values provided by methane absorption and the very low log(g) values provided by both methane absorption and CO overtones. The estimated metallicity is mostly solar, except for the fit considering the CO absorption forest, which yielded a supersolar value at the grid's edge. However, caution should be exercised when interpreting this solar value, as it appears that the fits converged to a single node of the grid, possibly due to a bias in the interpolation step caused by small inhomogeneities in the grid. Two modes of C/O ratios are identified: one supersolar when carbon monoxide is considered in the fits, and one subsolar when methane and potassium are used. \texttt{Exo-REM} demonstrated successful reproduction of each spectral feature with a relatively low \chr compared to the other models, except for the potassium absorption lines where the depth is not fully reached.}

\subsubsection{\texttt{Sonora}}
\textcolor{black}{The \Teff estimates obtained from \texttt{Sonora} range from 1170 to 1401~K, with a deviant value of 1837~K when considering the CH$_{4}$ feature. These results are consistent with the two previous models and with Section \ref{sec:windows_fit}. In contrast, we observed high surface gravities, with values exceeding 4.5~dex, which are inconsistent with the age of the system. Additionally, a supersolar metallicity is found, except when fitting using CH$_{4}$. Similar to the adiabatic index $\gamma$ explored by \texttt{ATMO}, the sedimentation factor \fsed also varies across the size of the grid. \texttt{Sonora} is capable of reproducing the different spectral features and generally exhibits the second lowest \chr compared to the other grids.}

\subsubsection{\texttt{BT-Settl}}
\textcolor{black}{This model estimates the \Teff to be between 1426 and 1718~K, with a low surface gravity of $<$ 4.5~dex, which is coherent with the spectral type and age of the object, respectively. Despite its ability to reproduce each spectral feature, the calculated \chr is generally higher than that of the three previous models we discussed, indicating lower performance.}

\subsubsection{\texttt{DRIFT-PHOENIX}}
\begin{figure*}[!t]
\centering
\includegraphics[width=1.0\hsize]{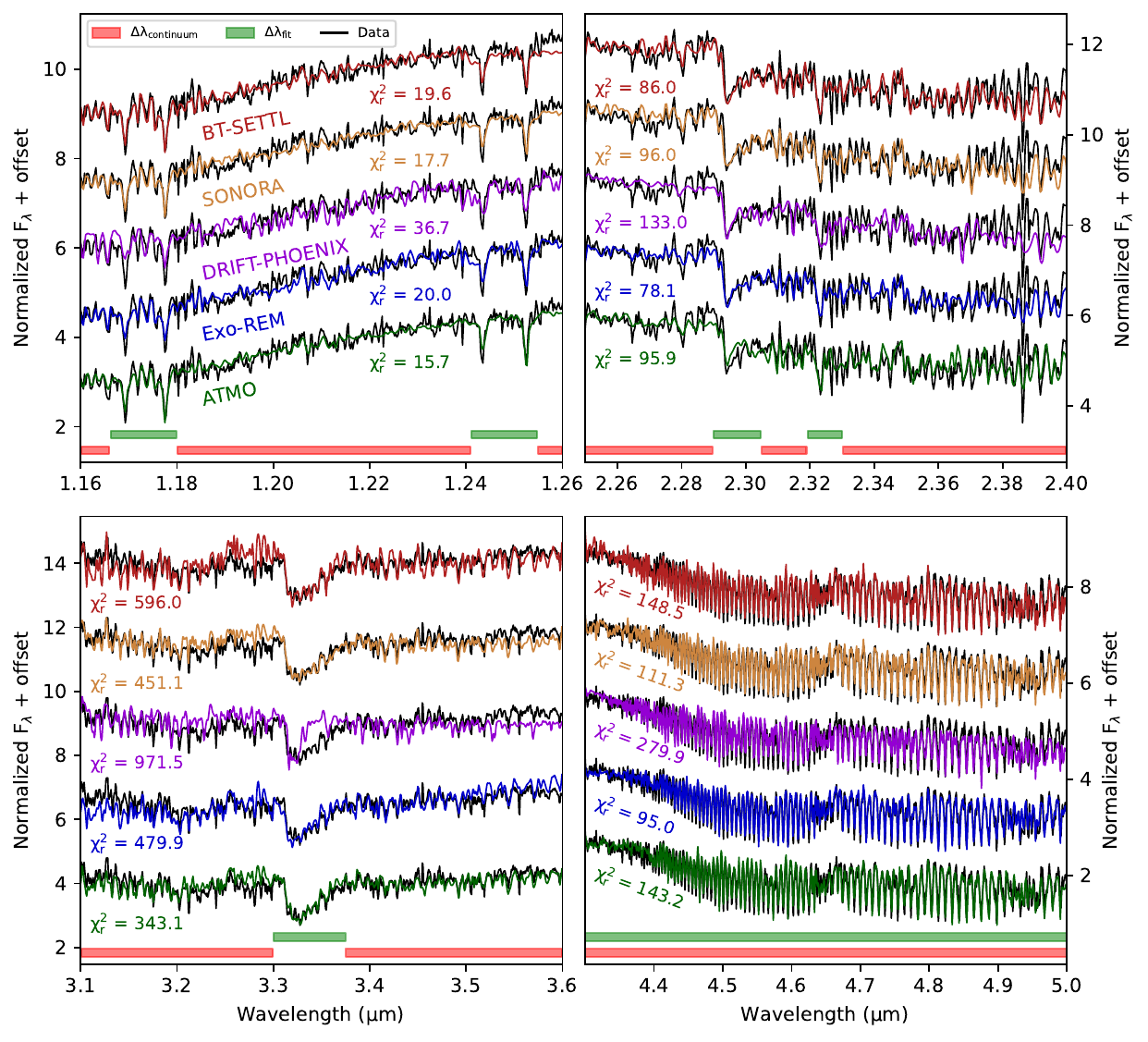}
    \caption{Comparison between the data and the interpolated synthetic spectra generated with the set of parameters that maximized the likelihood for each model, and for each spectral feature considered. The red area represents the wavelength coverage used to estimate the \Teff and the dilution factor $\rm C_{K}$, while the green area represents the wavelengths used to fit the spectral features with \Teff and $\rm C_{K}$ fixed. The calculated $\rm \chi^{2}_{r}$ values for each fit are provided, allowing for a comparison of each model's ability to reproduce the data.}
    \label{fig:high_res_wind}
\end{figure*}

\begin{figure*}[hbtp!]
\vspace{0.5cm}
\centering
\includegraphics[width=0.85\hsize]{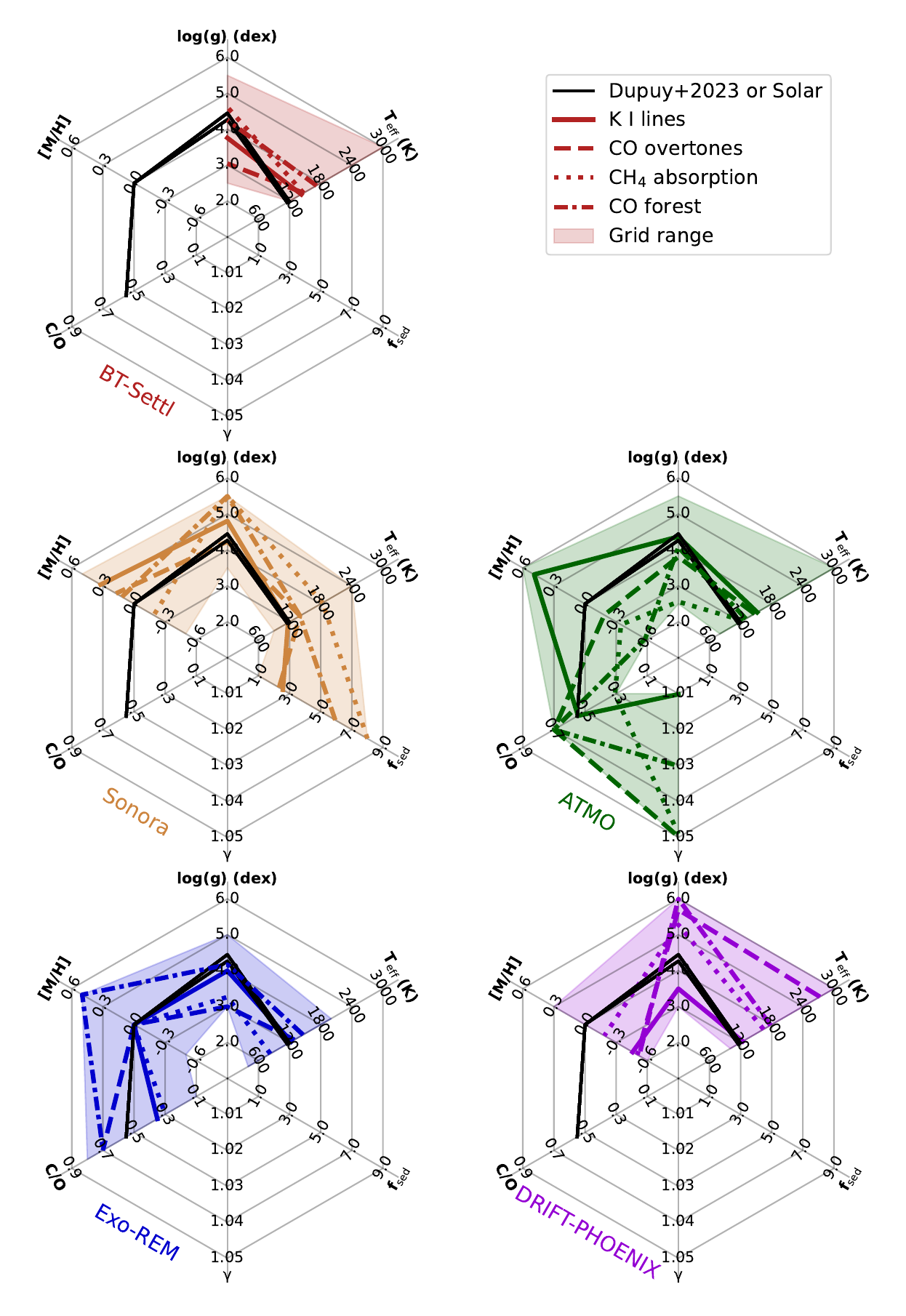}
    \caption{Parameter estimates provided by each model. The results obtained using each spectral feature are represented by different line formats. The colored areas represent the parameter spaces explored by each grid. The black lines indicate the estimates of \Teff and log(g) from \cite{Dupuy23} for the two mass scenarios, as well as the solar values of [M/H] and C/O. }
    \label{fig:high_res}
\end{figure*}

\textcolor{black}{As shown in Figure \ref{fig:high_res_wind} and by its high \chr values, this particular version of \texttt{DRIFT-PHOENIX} was unsuccessful in accurately reproducing the spectral information at medium resolution. Consequently, making a robust estimate of atmospheric parameters becomes challenging.}


\section{Discussion}
\label{sec:discussion}
\textcolor{black}{In Sections \ref{sec:windows_fit} and \ref{sec:features_fit}, we have presented two different approaches to leverage the spectral information observable from imaged exoplanets with JWST, aiming to estimate their atmospheric properties. The results of this study indicate a dispersion of these estimated parameters based on the wavelength coverage used for the fit, the considered spectral features, and the adopted model. We have proposed a method to incorporate this dispersion into a new error bar, attempting to provide a robust estimate for each parameter. This section is dedicated to discussing these results, investigating the origin of the dispersion, and analyzing the detected silicate absorption at 10~\mic.}

\subsection{The bolometric luminosity and the mass}

\begin{figure*}[t!]
\centering
\includegraphics[width=0.49\hsize]{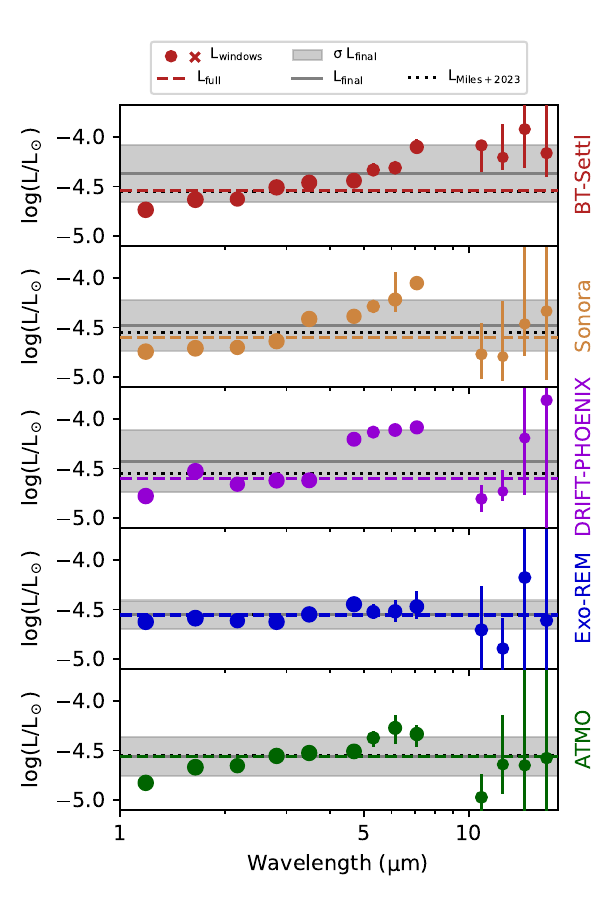}
\includegraphics[width=0.49\hsize]{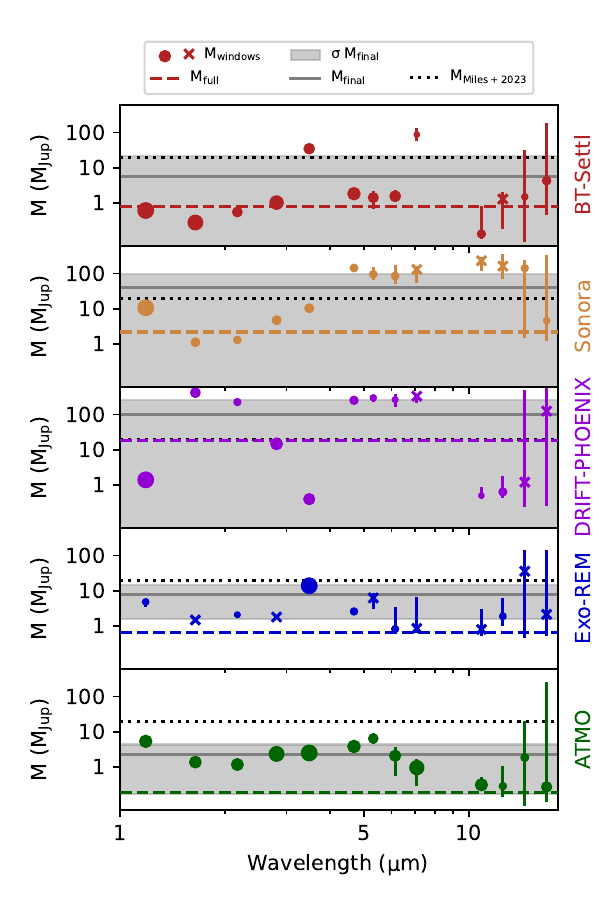}
    \caption{Same figure as Figure \ref{fig:teff_logg_compare}, but for the bolometric luminosity $\rm L_{bol}$ (left panel) and the mass M (right panel). These parameters are estimated from the Stefan-Boltzmann law and the gravitational law, respectively.}
    \label{fig:lum_m_compare}
\end{figure*}

\textcolor{black}{This JWST spectrum covers 98\% of the bolometric flux of VHS~1256~b. By filling the remaining 2\% with a model and integrating this full SED \cite{Miles22} calculated a log(L$\rm _{bol}$/L$\rm _{\odot}$) of -4.550~$\pm$~0.009, which was consistent with previous estimates provided by \cite{Hoch22} and \cite{Petrus23}. In our study, we derived the bolometric luminosity directly from our estimates of the \Teff and the radius R. For each spectral window defined in Section \ref{sec:windows_fit_meth}, we calculated L$\rm _{bol}$ using the Stefan-Boltzmann law: $\rm L_{bol} = 4\pi\sigma \times R^{2} \times T_{eff}^{4}$, where $\rm \sigma$ is the Stefan-Boltzmann constant. These results are visualized in Figure \ref{fig:lum_m_compare}, and as expected, we observe dispersion across the different spectral windows, influenced by the dispersion observed for the \Teff and the radius estimates. The fits using the full SED yielded bolometric luminosity in good agreement with the estimates of \cite{Miles22} for each model but with a very small relative error (less than 0.03\%). Using the same method described in Section \ref{sec:res_part1_final}, we calculated robust values of L$\rm _{bol}$ for each model and reported them in Table \ref{tab:final_part1_extragrid}. \texttt{Exo-REM} and \texttt{ATMO} are the two models that provide the most stable L$\rm _{bol}$ over wavelength coverage.}

\textcolor{black}{We also calculated the mass from our estimates of the log(g) and the radius, applying the gravitational law:}

\begin{equation}
    \rm m~g = \frac{G~M~m}{R^{2}} \longleftrightarrow  M = \frac{g~R^{2}}{G}
\end{equation}

\textcolor{black}{where $G$ is the gravitational constant. As with the bolometric luminosity, we also observe dispersion in the mass estimates (Figure \ref{fig:lum_m_compare}). We take this dispersion into account to provide the final values reported in Table \ref{tab:final_part1_extragrid}. Interestingly, for the grids \texttt{BT-Settl}, \texttt{Exo-REM}, and \texttt{ATMO}, the mass values obtained from each spectral window appear to be too low and unphysical ($<5$~\Mjup) when compared to the previous estimate made by \cite{Miles22}. They considered the age of the system, the measured L$\rm _{bol}$, and the hybrid cloudy evolutionary model of \cite{Saumon08} to estimate the mass of VHS~1256~b to be lower than 20~\Mjup with two potential solutions: one at 12~$\pm$~0.5~\Mjup and another at 16~$\pm$~2~\Mjup. Nonetheless, this comparison must be approached with caution due to the inherent uncertainties within evolutionary models, encompassing factors such as initial entropy variations and the uncertainty in age determinations, including the potential for age discrepancies between the star and its companion. The trend of log(g) as shown in Figure \ref{fig:teff_logg_compare} seems to impact the \texttt{Sonora} grid, indicating a higher mass when using MIRI data and a lower mass when using NIRSpec data. Lastly, the substantial dispersion in log(g) obtained with the \texttt{DRIFT-PHOENIX} grid results in a wide range of mass estimates.}

\subsection{The origins of the dispersion}
\label{dis:origins}
\textcolor{black}{The results presented in Sections \ref{sec:windows_fit} and \ref{sec:features_fit} have revealed that the estimates of atmospheric properties for VHS~1256~b, derived using the forward modelling approach, exhibit significant dependence on several factors. These include the wavelength coverage, the specific model used, and the type of information considered for the fit (large spectral windows or specific spectral features). The dispersion observed in Figures \ref{fig:teff_logg_compare}-\ref{fig:radius_compare} can be attributed to three main sources: the object's characteristics, the data reduction, and the models' performances and properties (chemical abundances, physics considered).}

\textcolor{black}{Firstly, VHS~1256~b has been identified as the most variable substellar object by \cite{Bowler20} and \cite{Zhou22}, showing a luminosity variation of over 20\% between 1.1 and 1.7~\mic over an 8-hour period and $\sim$5.8\% at 4.5~\mic \citep{Zhou20}. This variability has been attributed to an inhomogeneous cloud cover. A spectroscopic time follow-up of this object is underway to analyze its cloud and atmospheric structure, and the results will be presented in dedicated papers. Using the three-sinusoid model proposed by \cite{Zhou22}, \cite{Miles22} conducted a simulation to estimate the expected variability during the 4-hour time acquisition of the JWST data used in this study. They estimated a maximum variability of less than 5\% for 1~\mic$< \uplambda< $ 3~\mic, about 1.5\% for 3~\mic$< \uplambda< $ 5~\mic, and negligible variability for $\uplambda > $ 5~\mic. However, Figure \ref{fig:lum_m_compare} reveals an unexpected increase in the best fit bolometric luminosity as a function of wavelength for fitted spectral channels between 1 and 7~\mic, with approximate percentage changes of 330\%, 390\%, 400\%, 40\%, and 210\% for the grids \texttt{BT-Settl}, \texttt{Sonora}, \texttt{DRIFT-PHOENIX}, \texttt{Exo-REM}, and \texttt{ATMO}, respectively, which cannot be attributed to the variability of VHS~1256~b.}

\textcolor{black}{An alternative explanation for this luminosity dispersion might be attributed to the extraction methodology. The dataset employed in this analysis comprises distinct channels observed independently (3 for NIRSpec and 9 for MIRI). As the spectral coverage provided by these channels is employed to define the spectral windows used in the fitting process in Section \ref{sec:windows_fit}, inaccuracies in flux calibration during extraction could potentially influence the inferred luminosity. However, it appears improbable, given that the same discrepancy is discernible between the windows of each three couples of windows defined from the three NIRSpec channels, implying consistency within the extraction process (see Table \ref{tab:spec_wind_nirspec}). Furthermore, the magnitude of the dispersion varies across models, indicating a model-dependent interpretation.}

\textcolor{black}{Previous studies have invoked systematic errors within models of atmospheres to explain biases in estimating atmospheric parameters for early-type and hotter objects. \cite{Petrus20} attributed them to deficiencies in dust grain modelling, which were partially mitigated by introducing interstellar extinction ($\rm A_{V}$) as an additional free parameter in the fitting process. This approach was further employed by \cite{Hurt23} in the characterization of 90 late-M and L dwarfs. They confirmed that the fit quality significantly improved when incorporating an extra source of extinction. Notably, no interstellar extinction was detected for these observed targets. While both \cite{Petrus20} and \cite{Hurt23} focused on data within the wavelength range $\uplambda <$ 2.5~\mic, we extended this concept to a broader wavelength range. We incorporated the interstellar extinction law as defined by \cite{Draine03} as a free parameter in the fitting process, using our five different models.}

\textcolor{black}{The results, as presented in Appendix \ref{App:av}, demonstrate that exploring the impact of an interstellar extinction leads to an enhancement in the quality of the fits for the models \texttt{BT-Settl}, \texttt{Exo-REM}, and \texttt{ATMO}, evident through reduced \chr values. This improvement is not observed in the case of the \texttt{DRIFT-PHOENIX} and \texttt{Sonora}, where the calculated \chr values across the entire SED are higher when considering $\rm A_{V}$. This phenomenon can be explained by the significant heterogeneity inherent in the dataset concerning spectral resolution and SNR, as illustrated in the top-right panel of Figure \ref{fig:fits}. Notably, focusing exclusively on the channel G395HF/290LP of NIRSpec (2.87 to 5.27~\mic), which holds the greatest sensitivity in the likelihood computation, yields a reduction in \chr values ($\rm \chi^{2}_{r}$~=~126 and $\rm \chi^{2}_{r, A_{V}}$~=~116 for \texttt{DRIFT-PHOENIX}; and $\rm \chi^{2}_{r}$~=~297 and $\rm \chi^{2}_{r, A_{V}}$~=~148 for \texttt{Sonora}). With each model, substantial deviations persist when considering A$\rm _{V}$, particularly at $\uplambda >$ 5~\mic, underscoring that the considered extinction law, derived from a population of dust grains within the interstellar medium, fails to accurately capture the dust and cloud characteristics that shape the atmosphere of VHS~1256~b. We have also observed a negative value of $\rm A_{V}$ when the model \texttt{Exo-REM} is considered, indicating that this model is excessively red. This underscores the critical significance of advancing the development of extinction laws specifically tailored for characterizing atmospheric dust and haze.}

\textcolor{black}{Moreover, the luminosity depicted in Figure \ref{fig:lum_m_compare} is calculated based on the estimated values of \Teff and radius considering the Stefan-Boltzmann law. Consequently, the dispersion of the luminosity within the 1 to 7~\mic range directly correlates with variations in these two parameters. Upon comparing the dispersions of \Teff and radius illustrated in Figures \ref{fig:teff_logg_compare} and \ref{fig:radius_compare}, respectively, it becomes apparent that the \Teff estimates remain relatively consistent across all models, except for \texttt{Sonora}. Conversely, the radius estimates show a general increase across all models, except for \texttt{Sonora}. One plausible explanation for this behavior is the potential variance in cloud cover between the \texttt{Sonora} model and the remaining models, attributed to the parameter \fsed, which governs the sedimentation process of the diverse clouds generated \cite{Luna21}. Indeed, it is expected that different cloud coverage would strongly impact the shape of the SED at low spectral resolution (Whiteford et al. in prep) but also should be critical in the reproduction of the spectral information at higher resolution (see Section \ref{sec:features_fit}) and consequently in the robust estimate of chemical abundances. Moreover, the cloudless model \texttt{ATMO}, grounded in chemical disequilibrium physics, offers good performances in reproducing the data. We suggest that this cloudless approach should be integrated with cloud models, as a synergistic interplay between the two factors is plausible and can yield combined effects. Indeed, in the case of VHS~1256~b, the presence of clouds in the photosphere is directly detected from the spectra, as evidenced by the silicate absorption. This implies that a fully cloud-free model is evidently not self-consistent in reproducing the atmosphere of this kind of object.}

\textcolor{black}{Lastly, it's important not to disregard the potential influence of the spectral covariance and the choice of Bayesian priors which are known to impact the estimation of atmospheric parameters \citep{Greco16}. In this work, we've opted for a conservative approach by defining flat priors constrained by the size of the grids considered. Some inversion codes, like \texttt{STARFISH} \citep{Czekala15}, have integrated the impact of an error covariance matrix into likelihood calculations. This code also quantifies the influence of systematic errors in models using a Markov Chain Monte Carlo framework. In our future work, we plan to implement and test these functionalities while considering nested sampling in our code \texttt{ForMoSA} (Ravet et al., 2024, in preparation).}

\subsection{Exploitation of the silicate absorption}
\label{section:silicate}
\textcolor{black}{JWST's data provided the first detection of silicate absorption in the atmosphere of a low-mass companion \citep{Miles22}, directly revealing the presence of clouds in the atmosphere of VHS~1256~b. This feature had already been detected in the atmosphere of isolated objects, in particular at the L/T transition (e.g. \citealt{Cushing06}, \citealt{Looper08}, \citealt{Suarez22}). At these temperatures ($\sim$ 1200-1700~K) clouds of forsterite (Mg$_{2}$SiO$_{4}$), enstatite (MgSiO$_{3}$), and  quartz (SiO$_{2}$) can condense following non-equilibrium chemistry \citep{Lodders02, Helling06}. This has been detected empirically using atmospheric retrieval methods \citep{Burningham21, Vos23} and will be explored in the retrieval analysis of the dataset of VHS~1256~b used in this paper (Whiteford et al., in prep.). In their work, \cite{Suarez22} used a library of low-resolution spectra (\Rlam $\sim$ 90), obtained with Spitzer, to define a spectral index in order to investigate the depth of the silicate absorption as a function of the spectral type. Their library was composed of $\sim$ 110 field and young isolated planetary-mass objects covering a wide range of spectral types from M5 to T9. Their method was based on multiple linear interpolations to estimate the pseudo-continuum and the absorption depth at the position of the feature (at 9~\mic). The spectral index was defined as the ratio of both fluxes. Thanks to the diversity of their spectral library they identified positive correlations between their silicate index and the near-infrared color excess, and between their silicate index and the photometric variability, proving that silicate clouds had a great impact on the atmospheric properties of planetary-mass objects. The silicate index definition was improved in \cite{Suarez23b} considering the silicate absorption extends up to 13~\mic in relatively young objects, like VHS~1256~b, and a better approach to estimate the continuum. This led to the main conclusion that the silicate absorption is sensitive to surface gravity. It is redder, broader, and more asymmetric in low-surface gravity dwarfs. This work was followed by \cite{Suarez23} who identified, in the same Spitzer library, a dependence between the depth of the silicate absorption and the viewing geometry of the target, confirming that equatorial latitudes are cloudier. In the case of VHS~1256~b, which is seen equator-on (90$^{+0}_{-28}$~deg; \citealt{Zhou20}), they find a strong silicate absorption and reddening. However, the inability to date to observe the silicate absorption between 8.0 and 12.0~\mic at medium resolution (\Rlam~$>$~100) has restricted the comparison of this molecular absorption's properties (depth, shape, etc.) with atmospheric model predictions. This limitation has slowed down the development of these models, as they currently cannot replicate this feature. Consequently, our comprehension of the physical and chemical mechanisms underlying its appearance, notably its chemical composition, remains hindered. In this section, we propose a new method, based on the forward modelling analysis to define a new silicate index.}

\subsubsection{Method}
\textcolor{black}{To avoid the region where the models fail to reproduce the silicate absorption, we defined two windows on the left side (6.0 to 8.0~\mic) and the right side (12.0 to 14.0~\mic) of the feature. The data were fitted using \texttt{ForMoSA}, considering only the wavelengths covered by these two windows. The parameters estimated from this fit were then used to estimate the pseudo-continuum between 8 and 12~\mic. We performed this fit for each model defined in Section \ref{sec:mod}, as well as for each isolated object in the Spitzer library and for the JWST data of VHS~1256~b. We limited the Spitzer library to objects with spectral types between M6 and T8. The final sample consisted of 12 young objects (age $<$400 Myr) and 93 field objects. The index $I_{Si}$ was calculated as the equivalent width of the silicate absorption using the formula:}

\begin{equation}
\label{eq:sili_index}
    I_{Si} = EW(Si) = \int_{\lambda} \left(1 - \frac{F_{\lambda}}{C_{\lambda}}\right) d\lambda
\end{equation}

\begin{figure}[!t]
\centering
\includegraphics[width=\hsize]{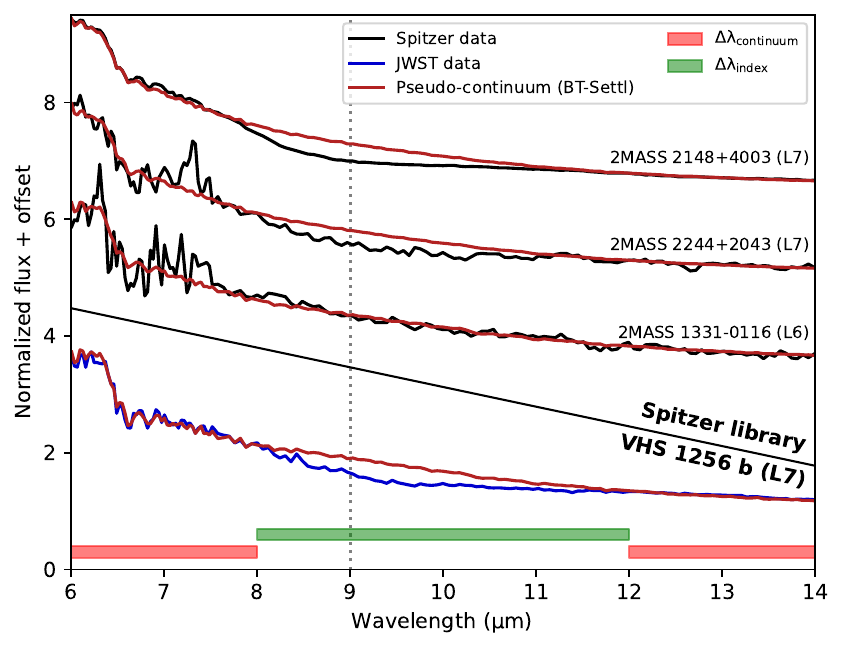}
    \caption{Comparison between three examples of silicate absorption identified in the Spitzer library and the detection in JWST data of VHS~1256~b. The spectral resolution of the JWST data has been reduced to match that of Spitzer for a fair comparison. The red area represents the spectral window used to estimate the pseudo-continuum, while the green area indicates the spectral window used to calculate the spectral index. In this figure, we specifically present the case of \texttt{BT-Settl} as an illustrative example.}
    \label{fig:method_sili}
\end{figure}

\textcolor{black}{where $C_{\lambda}$ and $F_{\lambda}$ are the fluxes of the estimated pseudo-continuum and the data, respectively, at the wavelength $\lambda$, with $\lambda \in$ [8.0, 12.0]~\mic. The error bar on $I_{Si}$ is estimated by the calculation of $I_{Si, min}$ and $I_{Si, max}$, considering $F - F_{err}$ and $F + F_{err}$, respectively, where $F_{err}$ is the error of the data. This method is illustrated in Figure \ref{fig:method_sili}.}

\subsubsection{Results}

\begin{figure*}[!t]
\centering
\includegraphics[width=0.8\hsize]{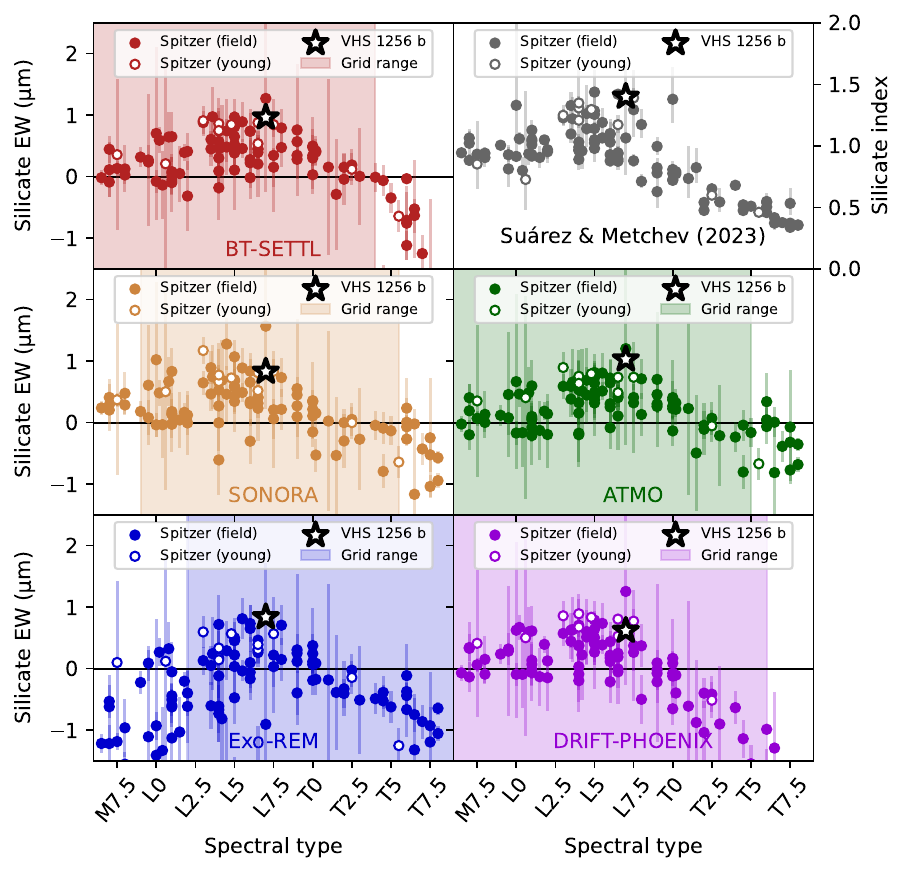}
    \caption{The silicate index as a function of spectral type for the objects in the Spitzer library and VHS~1256~b. The top-right panel displays the results obtained using the method defined by \cite{Suarez23b} for comparison. The other panels present the results obtained in this study, where the silicate index is calculated as an equivalent width. The open circles and filled circles represent young and field objects, respectively, and the star represents the estimate for VHS~1256~b. The colored areas indicate the spectral type ranges covered by the five different models used in the analysis.}
    \label{fig:results_sili}
\end{figure*}

\textcolor{black}{In their work, \cite{Suarez22} found that the silicate index was higher on average between the spectral types L4–L6 compared to colder and warmer objects in their library. They interpreted this phenomenon as the result of Si being unable to condense into clouds for \Teff $>$ 2000 K and the sedimentation of clouds below the photosphere after the L-T transition. Figure \ref{fig:results_sili} displays $I_{Si}$ as a function of spectral type for each object in the Spitzer library, including VHS~1256~b. The top-right panel reproduces the silicate index as defined by \cite{Suarez23b}, enabling a direct comparison between the two approaches. Similar results were obtained for each model used. An $I_{Si} \sim$ 0 is calculated for M-type objects, indicating a lack of significant silicate absorption in this spectral range. Additionally, the detection of CH$\rm _{4}$ after the L-T transition, as well as the potential contribution of the NH$\rm _{3}$ absorption at 10.5~\mic, which is present in the spectra of T dwarfs, may lead to negative values for the estimated equivalent widths. However, it should be noted that the validity of the results from the five models is limited to the spectral types they cover, as indicated by the colored area. Furthermore, we observe that the silicate index is generally higher for young objects compared to field objects, as recently highlighted by \cite{Suarez23b}. This is likely caused by the higher surface gravity of older objects, resulting in their contraction, which increases the sedimentation rate of the silicate clouds below the photosphere.}

\textcolor{black}{The silicate index calculated for VHS~1256~b, following the equivalent width procedure, aligns with the observed trends. The L7 spectral type and young age of this object favor the presence of a silicate absorption feature, resulting in a high value of $I_{Si}$ for each model. As observed in Figure \ref{fig:method_sili}, the absorption feature of VHS~1256~b seems to be redder and broader compared to the absorption detected in the spectrum of other objects (e.g., 2MASS~2148+4003). The reason for this distinct shape needs to be investigated, but it could be attributed to complex silicate clouds with varying abundances of different silicate elements. Indeed, \cite{Cushing06} showed that forsterite (Mg$_{2}$SiO$_{4}$) was redder than enstatite (MgSiO$_{3}$). The in-depth analysis of the silicate feature is a relatively new research direction that is currently in progress (\citealt{Burningham21}, \citealt{Suarez23}, \citealt{Suarez23b}). With this analysis, we have demonstrated that the current generation of atmospheric models can be utilized to accurately estimate the pseudo-continuum at the position of this feature, taking advantage of their inability to reproduce the silicate absorption feature. This allows for robust quantification of the equivalent width and facilitates discussions on the spectral type and age of observed objects.}

\subsection{Current and future improvements in models}
This work highlighted the current limitations of forward modelling in characterizing the atmospheres of low-mass objects. One approach to overcome these limitations is to improve the quality of the existing generation of atmospheric models. As demonstrated in this study, it's crucial to compare the outcomes obtained by considering various model families. Therefore, simultaneous improvements are necessary for each model. Various enhancement points have been identified.

The physics governing the clouds (formation and evolution) requires improvement. Currently, the particle size distribution is defined by a single broad log-normal distribution, which may need adaptation because it suggests smaller particles at higher atmospheric levels. Incorporating microphysical models that present smaller particles in a 'nucleation mode' and larger particles in a 'growth/settling mode' could be an interesting solution (see \citealt{Helling06}). Having a cloud model with smaller particles at the right heights might aid in reproducing the $\sim$10~\mic silicate feature and potentially explain some of the observed interstellar reddening from optical to near-infrared wavelengths.

As demonstrated by \texttt{ATMO}, diabatic convection can replicate cloud effects in the atmosphere, substantially altering temperature structures, particularly at the L/T transition. However, the presence of clouds is now confirmed through the detection of silicate absorption. Simultaneous integration of disequilibrium and clouds marks a promising advancement, slated for implementation in \texttt{Sonora} and refinement in \texttt{Exo-REM}. Additionally, introducing physically motivated adjustments to temperature structures, including effects like mean molecular weight gradients not presently considered, could impact the pressure-temperature profile in the mid- and upper-atmosphere, influencing cloud physics and estimates of chemical abundances.

All the model families considered in this study are 1D, combining simulations of the vertical atmosphere structure with radiative transfer codes to generate synthetic spectra. VHS~1256~b exhibits significant variability, suggesting a complex atmosphere potentially marked by an inhomogeneous cloud cover. It's evident that 1D models alone won't suffice to capture the complexity of these objects existing in three dimensions. Currently, significant efforts are underway to interpret such variability by developing a new generation of 3D  models of atmosphere \citep{Showman20, Tan21b, Tan21a, Plummer22} that aim to map the surfaces of exoplanets observed by JWST and, in the near future, by the ELT.

Finally, the recent identification of isotopologue ratios as tracers of formation motivates the development of models that can explore their potential different values. Recently, isotopologues ($^{13}$C, $^{18}$O, and $^{17}$O) were identified in the atmosphere of VHS~1256~b through comparisons of JWST data with synthetic spectra generated within a retrieval framework \citep{Gandhi23}. It's important to investigate the capability of self-consistent models in detecting these chemical elements, estimating their abundance, and studying the impact of systematic errors in the models on these values. Expansion of grids is currently underway to address this new avenue of research.

\section{Summary and conclusion}
\label{sec:summary}
\textcolor{black}{The main objective of this study was to conduct a comprehensive analysis, employing the forward modelling approach, of the most detailed spectrum obtained to date for an imaged planetary-mass companion. The aim was to assess the capabilities and limitations of the current generation of five different self-consistent atmospheric models and determine the extent to which we can characterize the atmosphere of such objects.}

\textcolor{black}{In Section \ref{sec:windows_fit}, we presented an innovative method that took advantage of the extended wavelength coverage provided by JWST. By dividing the SED into multiple spectral windows, we explored the dispersion of parameter estimates across these windows and propagated this dispersion to determine new error bars for each atmospheric property. Our analysis revealed that relatively robust estimates could be obtained for the \Teff, log(g), and radius. However, we observed that the log(g) and radius tended to be under-estimated compared to the predictions of evolutionary models. This discrepancy resulted in underestimated values of the mass. On the other hand, the calculated luminosity, derived from \Teff and radius, was in good agreement with evolutionary models. Interestingly, we observed a curious trend of increasingly large fitted bolometric luminosities as a function of wavelength region fit, which may be attributed to a misrepresentation of dust, haze, and clouds in the models of atmospheres. Estimating chemical abundances ([M/H] and C/O), as well as parameters such as $\rm \gamma$ and \fsed, proved to be more challenging due to the significant dispersion observed along the SED. Furthermore, found that the choice of the model used for the fit had a significant impact on the final parameter values, underscoring the importance of careful model selection in interpreting atmospheric properties, confirming the results of \cite{Lueber23} regarding the estimation of log(g).}

\textcolor{black}{In Section \ref{sec:features_fit}, we employed a targeted approach to constrain the [M/H] and C/O ratios by focusing on specific molecular and atomic absorption features known to be sensitive to these chemical abundances. By optimizing the fitting strategy for each absorption feature, we obtained consistent estimates of the \Teff and log(g) with those derived in Section \ref{sec:windows_fit}. Regarding [M/H], different scenarios emerged depending on the choice of the atmospheric model. The \texttt{Sonora} model suggested a supersolar metallicity, while the \texttt{DRIFT-PHOENIX} and \texttt{ATMO} models indicated a subsolar metallicity. Lastly, the \texttt{Exo-REM} model tended to converge towards solar values. Moreover, obtaining a robust estimate of the C/O ratio proved to be challenging as well. The \texttt{ATMO} model suggested supersolar values, similar to the \texttt{Exo-REM} model when considering CO, and coherent with previous studies \citep{Hoch22, Petrus23}. Nevertheless, in both cases, the inversion reached the edge of the grid.}

\textcolor{black}{After considering various sources of dispersion in the estimation of atmospheric parameters, such as the high variability of VHS~1256~b and potential biases during data reduction, we turned our attention to systematic errors in the five different models used. Indeed, there is a noticeable discrepancy in accurately reproducing the complexity of the atmosphere of VHS~1256~b, encompassing factors such as clouds' chemical composition and dispersion, properties of dust and haze, as well as various physical processes (vertical mixing, sedimentation, chemical disequilibrium, etc.). These systematics inherently contribute to biases in the determination of atmospheric parameters. Identifying and rectifying them will constitute one of the main challenges of exoplanetary atmospheric characterization, in the coming decade.}

\textcolor{black}{To characterize the silicate absorption feature detected, we took advantage of the fact that its absorption band at $\sim$10~\mic is not reproduced by the models. This allowed us to estimate the pseudo-continuum between 8 and 12~\mic, which enabled the calculation of the equivalent width of the silicate absorption. We compared it to those obtained from the silicate absorption detected in the spectra of isolated low-mass objects observed at low resolution with Spitzer. Through this analysis, we were able to confirm the correlation between the depth of the silicate absorption and the spectral type, as previously identified by \cite{Suarez22}. Additionally, we identified a connection between the silicate absorption feature and the age of the objects.}

\textcolor{black}{All of these results have demonstrated the capability of models to partially reproduce the observed data. The use of JWST data has allowed the development of optimized inversion procedures, which take into account the systematics present in the models. By performing a multi-model strategy and propagating these systematics into the error bars of the estimated parameters, we ensure a more accurate and realistic characterization of the atmospheric properties and avoid any potential over-interpretation of the results. However, constraining the formation history of VHS~1256~b using its atmospheric properties estimated from the forward modelling approach remains challenging. A supersolar metallicity is usually used to quantify the solid material that is accreted during the formation and the C/O is considered as a tracer of the birth location of the planet (see \citealt{GRAVITY20}, \citealt{Petrus21}, \citealt{Hoch22}, \citealt{Palma23}), but with this study, we have shown that the estimates of these commonly used tracers of formation were strongly dependent on the model and the spectral information used for the fit. Moreover, \cite{Molliere22} have demonstrated that the relationship between chemical abundances and the formation mechanisms described by the models (e.g., \citealt{Oberg11}) is more complex than initially anticipated, and it is dependent on various intricate parameters, particularly in the case of formation within a disk. The chemical composition of the protoplanetary disk from which VHS~1256~b may have formed is unknown, but it needs to be estimated in order to establish reference values. An alternative approach is to estimate it from the host star. However, in our case, VHS~1256~(AB) is a short-period binary, which poses significant challenges in accurately determining its chemical abundances. Furthermore, \cite{Molliere22} have demonstrated the significant influence of various factors on the evolution of the disk structure, such as gap opening, winds, and self-shadowing. They have also highlighted the impact of pebble evaporation inside ice lines on the accretion of chemical elements (both in gas and solid phases) during the planet's formation. These processes can potentially affect the overall metallicity and the carbon-to-oxygen ratio (C/O). Lastly, the authors emphasize that the chemical composition estimated solely from spectroscopic data may not necessarily reflect the bulk composition of the planet but rather describes the photospheric composition. In order to truly connect the measured atmospheric composition and the original disk composition, it is crucial to gain a comprehensive understanding of the mechanisms that facilitate vertical chemical exchange between the upper and deeper layers of the planetary atmosphere.}

\textcolor{black}{Due to the presence of these unknown factors, it seems challenging to definitively determine the mode of formation of VHS~1256~b, and more generally of planetary mass objects, by exclusively exploring its atmosphere. Nevertheless, the development and systematic application of spectral inversion methods, such as the one presented in this study, to the populations of imaged planetary-mass objects (both star companions and isolated objects), will enable a "comparative" analysis of their properties, evolving into a "statistical" study over the course of the JWST mission and once ELTs become operational around 2030.}










\section{Acknowledgments}
This work is based on observations made with the NASA/ESA/CSA James Webb Space Telescope. We are truly grateful for the countless hours that thousands of people have devoted to the design, construction, and commissioning of JWST. This project was supported by a grant from STScI (JWST-ERS-01386) under NASA contract NAS5-03127. This work benefited from the 2022 Exoplanet Summer Program in the Other Worlds Laboratory (OWL) at the University of California, Santa Cruz, a program funded by the Heising-Simons Foundation. S. P. acknowledges the support of ANID, --Millennium Science Initiative Program-- Center Code NCN19\_171. This project received funding from the European Research Council (ERC) under the European Union’s Horizon 2020 research and innovation programme (COBREX; grant agreement No. 885593) and from the ANR project FRAME (ANR-20-CE31-0012). J. M. V. acknowledges support from a Royal Society - Science Foundation Ireland University Research Fellowship (URF$\backslash$1$\backslash$221932). S. M. is supported by a Royal Society University Research Fellowship (URF-R1-221669). M. B. received funding from the European Union’s Horizon 2020 research and innovation programme (AtLAST; grant agreement No. 951815). R. A. M. is supported by the National Science Foundation MPS-Ascend Postdoctoral Research Fellowship under Grant No.~2213312. E.G. acknowledges support from the Heising-Simons Foundation for this research. I.R. is supported by grant FJC2021-047860-I and PID2021-127289NB-I00 financed by MCIN/AEI /10.13039/501100011033 and the European Union NextGenerationEU/PRTR. A. Z. and S. P. acknowledges support from ANID -- Millennium Science Initiative Program -- Center Code NCN2021\_080. All of the data presented in this article were obtained from the Mikulski Archive for Space Telescopes (MAST) at the Space Telescope Science Institute. The specific observations analyzed can be accessed via \dataset[10.17909/ceq5-9g20]{https://doi.org/10.17909/ceq5-9g20}. 

%





\appendix


\section{Focus on the fit comparison}
\label{App:zoom}
\begin{figure*}[b!]
\centering
\includegraphics[width=1.0\hsize]{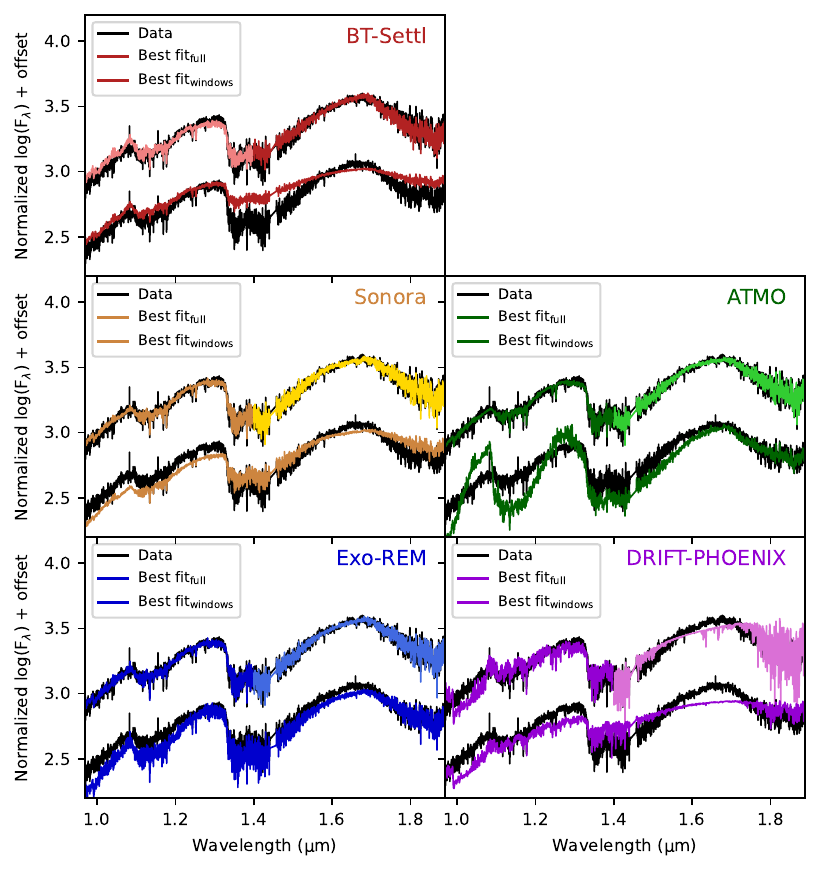}
    \caption{Same as Figure \ref{fig:fits}. A zoom has been performed to illustrate the differences between the models and the data with more details. In this figure, the zoom was chosen to represent the data acquired with the channel G140HF/100LP of NIRSpec, between 0.97 and 1.89~\mic.}
    \label{fig:zoom1}
\end{figure*}

\begin{figure*}[!t]
\centering
\includegraphics[width=1.0\hsize]{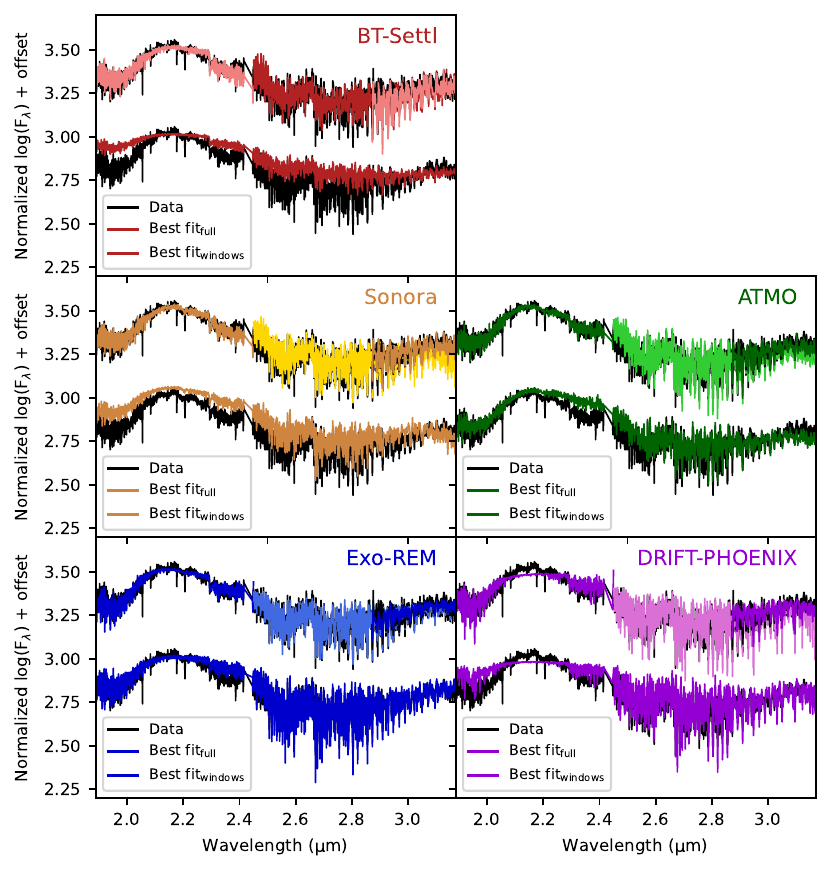}
    \caption{Same as Figure \ref{fig:zoom1} but with the zoom chosen to represent the data acquired with the channel G235HF/170LP of NIRSpec, between 1.89 and 3.17~\mic.}
    \label{fig:zoom2}
\end{figure*}

\begin{figure*}[!t]
\centering
\includegraphics[width=1.0\hsize]{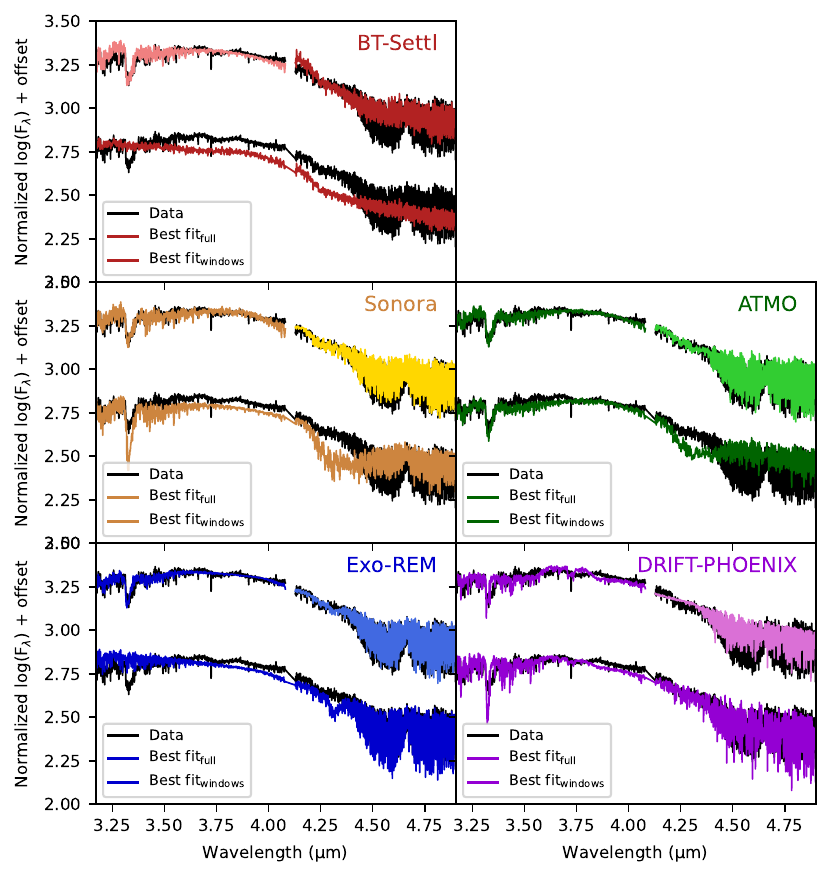}
    \caption{Same as Figure \ref{fig:zoom1} but with the zoom chosen to represent the data acquired with the channel G395HF/290LP of NIRSpec, between 3.17 and 5.27~\mic.}
    \label{fig:zoom3}
\end{figure*}

\begin{figure*}[!t]
\centering
\includegraphics[width=1.0\hsize]{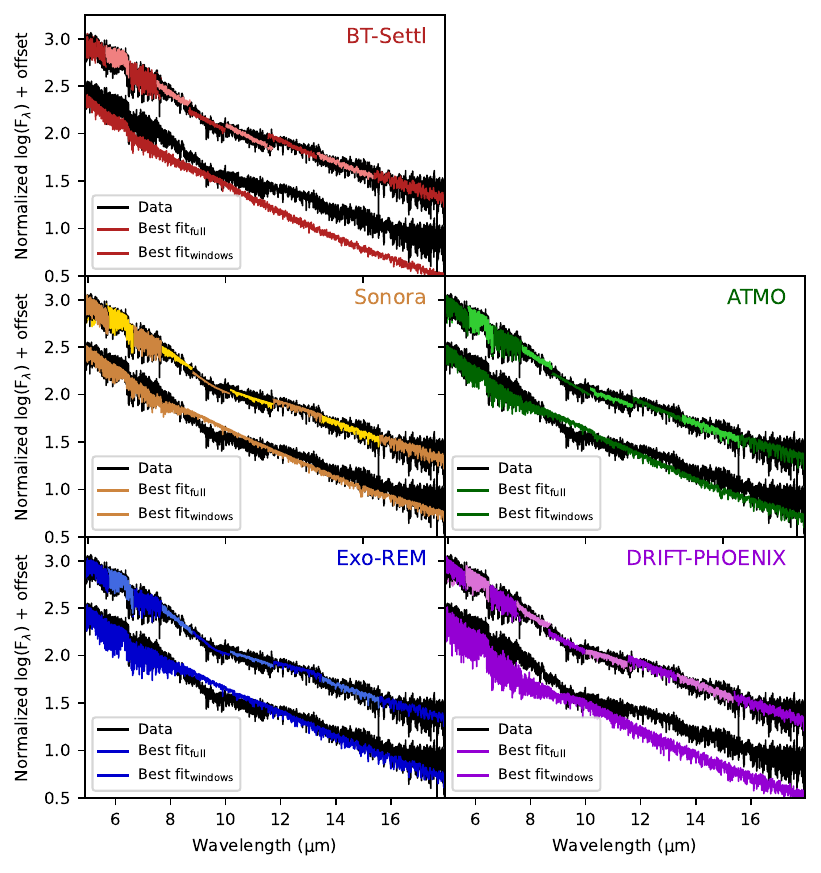}
    \caption{Same as Figure \ref{fig:zoom1} but with the zoom chosen to represent the data acquired with MIRI, between 4.90 and 17.98~\mic.}
    \label{fig:zoom4}
\end{figure*}

Figures \ref{fig:zoom1}-\ref{fig:zoom4} allow a comparison between the data and the best-fitting models with more details than in Figure \ref{fig:fits}.

\section{Interstellar extinction as a free parameter}
\label{App:av}
\begin{figure*}[!t]
\centering
\includegraphics[width=1.0\hsize]{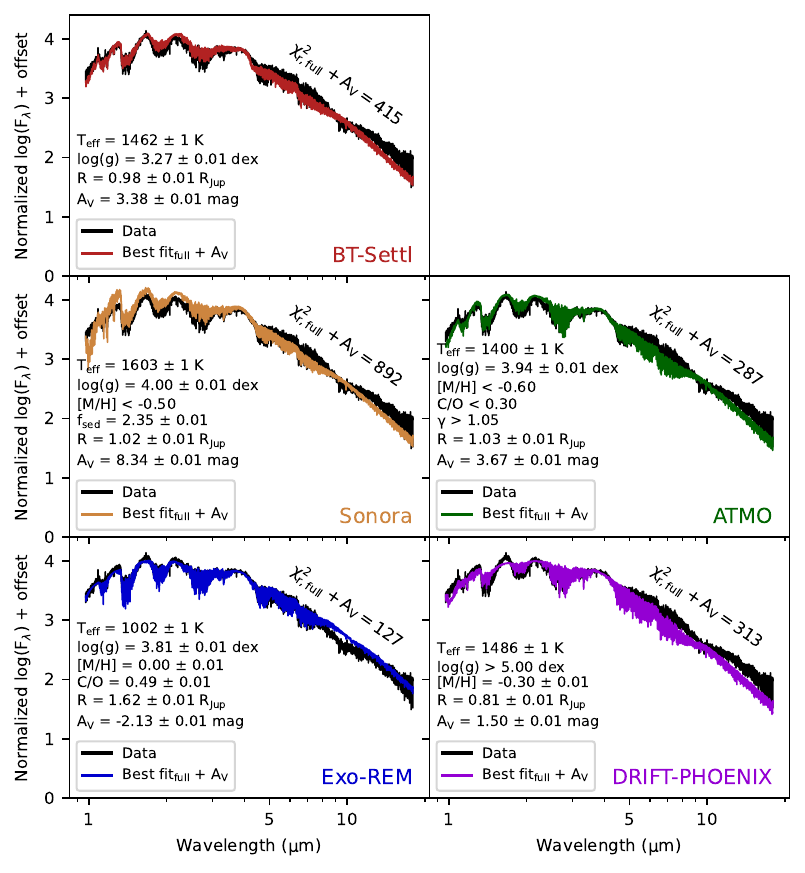}
    \caption{Same figure as Figure \ref{fig:fits}, but with the exploration of the interstellar extinction $\rm A_{V}$. The posterior of each other parameters are given.}
    \label{fig:fits_av}
\end{figure*}

Figure \ref{fig:fits_av} illustrates the best fitting model, achieved when integrating the interstellar extinction $\rm A_{V}$ as defined by \cite{Draine03} into the list of parameters explored by \texttt{ForMoSA}, following the concept introduced by \cite{Petrus20}. Notably, the \chr value is substantially diminished in comparison to cases where $\rm A_{V}$ is omitted (with the exception of \texttt{DRIFT-PHOENIX} and \texttt{Sonora}). However, despite these improvements, notable deviations persist in the reproduction of the observed data.

\newpage
 \,

\newpage
 \,

\newpage
 \,

\newpage
 \,

\newpage
 \,



\newpage
\bibliography{VHS1256b_JWST}
\bibliographystyle{aasjournal}





\end{document}